\documentclass[onecolumn]{IEEEtran}
\usepackage[mathscr]{eucal}
\usepackage[cmex10]{amsmath}
\usepackage{epsfig,epsf,psfrag}
\usepackage{amssymb,amsmath,amsthm,amsfonts,latexsym}
\usepackage{amsmath,graphicx,bm,xcolor,url,overpic}
\usepackage{fixltx2e}%ordering of single and double column floats
\usepackage{array}%array and tabular environments
\usepackage{verbatim}
\usepackage{bm}
\usepackage{algorithmic}
\usepackage{algorithm}
\usepackage{verbatim}
\usepackage{textcomp}
\usepackage{mathrsfs}
\usepackage{epstopdf}

\newcommand{\openone}{\leavevmode\hbox{\small1\normalsize\kern-.33em1}}

\catcode`~=11 \def\UrlSpecials{\do\~{\kern -.15em\lower .7ex\hbox{~}\kern .04em}} \catcode`~=13 

\allowdisplaybreaks[4]

\newcommand{\nn}{\nonumber}

\newcommand{\calA}{\mathcal{A}}
\newcommand{\calB}{\mathcal{B}}

\newcommand{\calD}{\mathcal{D}}
\newcommand{\calE}{\mathcal{E}}
\newcommand{\calF}{\mathcal{F}}

\newcommand{\calI}{\mathcal{I}}
\newcommand{\calJ}{\mathcal{J}}

\newcommand{\calL}{\mathcal{L}}

\newcommand{\calP}{\mathcal{P}}

\newcommand{\calS}{\mathcal{S}}
\newcommand{\calT}{\mathcal{T}}

\newcommand{\calW}{\mathcal{W}}
\newcommand{\calX}{\mathcal{X}}
\newcommand{\calY}{\mathcal{Y}}
\newcommand{\calZ}{\mathcal{Z}}

\newcommand{\ba}{\mathbf{a}}

\newcommand{\bi}{\mathbf{i}}

\newcommand{\bs}{\mathbf{s}}
\newcommand{\bS}{\mathbf{S}}

\newcommand{\bw}{\mathbf{w}}
\newcommand{\bW}{\mathbf{W}}
\newcommand{\bx}{\mathbf{x}}
\newcommand{\bX}{\mathbf{X}}

\newcommand{\rmb}{\mathrm{b}}

\newcommand{\rmc}{\mathrm{c}}

\newcommand{\rmd}{\mathrm{d}}

\newcommand{\rme}{\mathrm{e}}

\newcommand{\rmp}{\mathrm{p}}
\newcommand{\rmP}{\mathrm{P}}
\newcommand{\rmq}{\mathrm{q}}
\newcommand{\rmQ}{\mathrm{Q}}

\newcommand{\rmT}{\mathrm{T}}

\newcommand{\rmU}{\mathrm{U}}

\newcommand{\rmV}{\mathrm{V}}

\newcommand{\bbN}{\mathbb{N}}

\newcommand{\bbR}{\mathbb{R}}

\DeclareMathAlphabet{\mathbsf}{OT1}{cmss}{bx}{n}
\DeclareMathAlphabet{\mathssf}{OT1}{cmss}{m}{sl}% slanted sans serif

\DeclareSymbolFont{bsfletters}{OT1}{cmss}{bx}{n}  
\DeclareSymbolFont{ssfletters}{OT1}{cmss}{m}{n}
\DeclareMathSymbol{\bsfGamma}{0}{bsfletters}{'000}
\DeclareMathSymbol{\ssfGamma}{0}{ssfletters}{'000}
\DeclareMathSymbol{\bsfDelta}{0}{bsfletters}{'001}
\DeclareMathSymbol{\ssfDelta}{0}{ssfletters}{'001}
\DeclareMathSymbol{\bsfTheta}{0}{bsfletters}{'002}
\DeclareMathSymbol{\ssfTheta}{0}{ssfletters}{'002}
\DeclareMathSymbol{\bsfLambda}{0}{bsfletters}{'003}
\DeclareMathSymbol{\ssfLambda}{0}{ssfletters}{'003}
\DeclareMathSymbol{\bsfXi}{0}{bsfletters}{'004}
\DeclareMathSymbol{\ssfXi}{0}{ssfletters}{'004}
\DeclareMathSymbol{\bsfPi}{0}{bsfletters}{'005}
\DeclareMathSymbol{\ssfPi}{0}{ssfletters}{'005}
\DeclareMathSymbol{\bsfSigma}{0}{bsfletters}{'006}
\DeclareMathSymbol{\ssfSigma}{0}{ssfletters}{'006}
\DeclareMathSymbol{\bsfUpsilon}{0}{bsfletters}{'007}
\DeclareMathSymbol{\ssfUpsilon}{0}{ssfletters}{'007}
\DeclareMathSymbol{\bsfPhi}{0}{bsfletters}{'010}
\DeclareMathSymbol{\ssfPhi}{0}{ssfletters}{'010}
\DeclareMathSymbol{\bsfPsi}{0}{bsfletters}{'011}
\DeclareMathSymbol{\ssfPsi}{0}{ssfletters}{'011}
\DeclareMathSymbol{\bsfOmega}{0}{bsfletters}{'012}
\DeclareMathSymbol{\ssfOmega}{0}{ssfletters}{'012}

\newcommand{\tilM}{\tilde{M}}

\newcommand{\hats}{\hat{s}}

\newcommand{\hatw}{\hat{w}}

\newcommand{\hatx}{\hat{x}}

\newcommand{\tilx}{\tilde{x}}

\newtheorem{theorem}{Theorem}

\newtheorem{definition}{Definition}

\usepackage[utf8]{inputenc} % allow utf-8 input
\usepackage[T1]{fontenc}    % use 8-bit T1 fonts
\usepackage{url,bbm,balance}            % simple URL typesetting
\usepackage{booktabs}       % professional-quality tables
\usepackage{amsfonts,algorithm,algorithmic}       % blackboard math symbols
\usepackage{nicefrac}       % compact symbols for 1/2, etc.
\usepackage{microtype}      % microtypography

\usepackage{cite}
\usepackage{subfigure,transparent,color,graphicx} %used to insert pictures

\allowdisplaybreaks[2]
\newcommand{\bbo}{\mathbbm{1}}

\begin{document}

\title{Resolution Limits of Non-Adaptive 20 Questions Search for Multiple Targets}
\author{Lin Zhou, Lin Bai and Alfred O. Hero\\

\thanks{Copyright (c) 2017 IEEE. Personal use of this material is permitted. However, permission to use this material for any other purposes must be obtained from the IEEE by sending a request to pubs-permissions@ieee.org.}
}
\maketitle

\begin{abstract}
We study the problem of simultaneous search for multiple targets over a multidimensional unit cube and derive fundamental resolution limits of non-adaptive querying procedures using the 20 questions estimation framework. The performance criterion that we consider is the achievable resolution, which is defined as the maximal $L_\infty$ norm between the location vector and its estimated version where the maximization is over all target location vectors. The fundamental resolution limit is defined as the minimal achievable resolution of any non-adaptive query procedure, where each query has binary yes/no answers. We drive non-asymptotic and second-order asymptotic bounds on the minimal achievable resolution, using tools from finite blocklength information theory. Specifically, in the achievability part, we relate the 20 questions problem to data transmission over a multiple access channel, use the information spectrum method by Han and borrow results from finite blocklength analysis for random access channel coding. In the converse part, we relate the 20 questions problem to data transmission over a point-to-point channel and adapt finite blocklength converse results for channel coding. Our results extend the purely first-order asymptotic analyses of Kaspi \emph{et al.} (ISIT 2015) for the one-dimensional case: we consider channels beyond the binary symmetric channel and derive non-asymptotic and second-order asymptotic bounds on the performance of optimal non-adaptive query procedures.
\end{abstract}

\begin{IEEEkeywords}
Second-order asymptotics, query-dependent noise, finite blocklength analysis, information spectrum method, random access channel coding
\end{IEEEkeywords}

\section{Introduction}
R\'enyi initiated the study of a random search problem in~\cite{renyi1961problem}, which was motivated by diverse applications including medical diagnosis, chemical analysis and searching for a root of an equation. The problem was later known as the 20 questions game for estimation. In this problem, there are two players: an oracle and a questioner. The oracle knows the realization of a target random variable, say $S\in[0,1]$. The goal of the questioner is to accurately estimate $S$ by posing as few queries as possible, where each query has binary yes/no answers. In \cite{renyi1961problem}, the behavior of the oracle is modeled by a memoryless noisy channel and the noisy responses are the outputs of this channel with true answers to the queries being the inputs. Of central interest is the determination of the minimum number of queries needed to accurately estimate the target random variable using noisy responses from the oracle under a particular performance criterion.

Jedynak \emph{et al.}~\cite{jedynak2012twenty} revived interest in this problem by adopting a minimum entropy criterion on the posterior distribution of the target state. Specifically, Jedynak \emph{et al.} proposed an optimal non-adaptive query procedure named the dyadic policy in addition to an optimal adaptive query procedure. A non-adaptive query procedure asks a fixed number of predetermined queries while an adaptive query procedure poses a random number of queries and each query can depend on previous queries and the received noisy responses. Readers can refer to \cite{jedynak2012twenty} for detailed explanation of the differences between adaptive and non-adaptive query procedures. The adaptive query procedure of Jedynak \emph{et al.} is based on the probabilistic bisection policy in \cite{horstein1963sequential} (see also \cite{shayevitz2011} for additional analysis for channel coding with feedback). Subsequently, the results of Jedynak were generalized to a collaborative case~\cite{tsiligkaridis2014collaborative}, a decentralized case~\cite{tsiligkaridis2015decentralized} and a multiple target case~\cite{rajan2015bayesian}.

Inspired by future research directions pointed out in \cite{jedynak2012twenty}, a sequence of recent works including \cite{chung2018unequal,chiu2019noisy,kaspi2018searching,zhou2019twentyq} considered the more natural performance criterion such as the $L_2$ norm or the $L_{\infty}$ norm of the estimation error; furthermore, a subset of these works, such as \cite{chiu2019noisy,zhou2019twentyq} considered query-dependent noise where the noisy channel that models the behavior of the oracle depends on the size of the query. Such a query-dependent model is applicable to target localization tasks in a sensor network where the noise can accumulate when collecting responses or in a human-in-the-loop estimation model where workers make errors with error probability that is a function of the query.

Most previous work focuses on the case of a single target with rare exceptions, e.g.,~\cite{rajan2015bayesian,kaspi2015searching}. As discussed in the pioneering work on multiple target search by Rajan \emph{et al.}~\cite{rajan2015bayesian}, the problem is motivated by diverse applications, ranging from localizing faces in pictures to finding quasars in astronomical data. To advance the understanding of fundamental limits of multiple target search, in this paper, we take a finite blocklength information theoretical look at the problem.

\subsection{Main Contributions}

We obtain bounds on the performance of non-adaptive query procedures for multiple target search over a finite dimensional unit cube with a query-dependent noise model (cf. Definition \ref{def:mdchannel}). We define the fundamental resolution limit $\delta^*(n,k,d,\varepsilon)$ as the $L_{\infty}$ norm of the minimal achievable estimation error (cf. Eq. \eqref{def:delta*}) of any non-adaptive query procedure with $n$ queries that search for $k$ targets over a $d$-dimensional unit cube with error probability of at most $\varepsilon$, where each query asks whether any target lies in a certain search region and has only binary yes/no answers. 

We provide non-asymptotic bounds in Theorems \ref{theorem:fbl:achievability} and \ref{theorem:fbl:converse}, and a second-order asymptotic bound in Theorem \ref{result:second:ktarget} on the fundamental limit $\delta^*(n,k,d,\varepsilon)$ (cf. \eqref{def:delta*}). In particular, our second-order asymptotic result in Theorem \ref{result:second:ktarget} provides approximations to the performance of any optimal non-adaptive query procedure with finitely many queries. These approximations are validated via numerical examples in Section \ref{sec:simulation}. To prove our non-asymptotic achievability bound in Theorem \ref{theorem:fbl:achievability}, we relate the query problem to data transmission over a multiple access channel (MAC), employ random coding arguments and use the information spectrum method in \cite[Lemma 7.10.1]{han2003information}. To prove our non-asymptotic converse bound in Theorem \ref{theorem:fbl:converse}, we relate the current problem to data transmission over a point-to-point channel and adapt the finite blocklength converse analysis for channel coding~\cite[Proposition 4.4]{TanBook}. Finally, to prove the second-order asymptotic result in Theorem \ref{result:second:ktarget}, we apply the Berry-Esseen theorem~\cite{berry1941accuracy,esseen1942liapounoff} to the derived non-asymptotic bounds and we use inequalities from random access channel coding~\cite[Lemmas 1 and 2]{yavas2018random} and results from finite blocklength information theory~\cite[Lemma 49]{polyanskiy2010thesis}.

As a corollary of our second-order asymptotic result, we establish a phase transition phenomenon of the minimal excess-resolution probability as a function of the resolution decay rate. In particular, when the number of queries increases, the excess-resolution probability of an optimal non-adaptive query procedure switches rapidly from zero to one as a function of the resolution decay rate, where the critical threshold is a function of the capacity of the query-dependent noisy channel (cf. the first remark of Theorem \ref{result:second:ktarget} and Figure \ref{illus_pt_ktarget}). The results obtained in \cite{zhou2019twentyq} for optimal non-adaptive query procedures follow as a special case of the current result with $k=1$. Finally, our results extend easily to the $L_2$ norm performance criterion, which is also bounded by the fundamental limit $\delta^*(n,k,d,\varepsilon)$.

\subsection{Related Works}
There is a vast amount of literature on the 20 questions problem, but very few address the problem of simultaneous search for multiple targets, e.g.,~\cite{rajan2015bayesian,kaspi2015searching}. The authors of \cite{rajan2015bayesian} considered the minimal posterior entropy criterion introduced in \cite{jedynak2012twenty}, which is different from the resolution criterion adopted in the current paper. Therefore, we compare our results with those established in \cite[Theorem \ref{theorem:fbl:achievability}]{kaspi2015searching}. The first difference is that we consider an arbitrary binary input query-dependent discrete memoryless channel (DMC) and allow the target to be any finite dimension $d$, while \cite{kaspi2015searching} was restricted to the query-dependent binary symmetric channel (BSC) with $d=1$\footnote{
The multiple target search analysis of \cite{kaspi2015searching} is also based on random coding arguments. This line of argument can be generalized to our more general setting for establishing achievability of first-order asymptotic error bounds.}. Secondly, we derive the second-order asymptotic result on the fundamental limit of non-adaptive query procedures, which provides benchmarks for the practical case of \emph{finitely} many queries,  while \cite{kaspi2015searching} only performed a first-order asymptotic analysis characterizing the resolution decay rate in the limit of an \emph{infinite} number of queries. Thirdly, we offer new insights beyond those of \cite{kaspi2015searching}, including establishing a phase transition phenomenon that demonstrates a sharp degradation of the excess-resolution probability in the noise variance as a function of the resolution decay rate. Finally, our proof techniques significantly differ from \cite{kaspi2015searching} where asymptotic Shannon techniques including the Fano's inequality were used. In contrast, we use the recently developed finite blocklength information theoretical tools of \cite{polyanskiy2010finite,TanBook,yavas2018random} and the information spectrum method~\cite[Lemma 7.10.1]{han2003information}.

While the contributions of this paper go well beyond our previous work~\cite{zhou2019twentyq} and random access channel coding~\cite{yavas2018random}, some of our proof techniques are adapted from them. In~\cite{zhou2019twentyq}, we derived fundamental limits on 20 questions estimation for a single target using non-adaptive query procedures. In this paper, we generalize these results to simultaneous search for multiple targets. Such a generalization is not straightforward and the methods of the proof are very different from~\cite{zhou2019twentyq}. The achievability proof techniques that we use for multiple targets are inspired by multiple access channel coding~\cite{han2003information} and random access channel coding~\cite{yavas2018random}, while the main proof techniques for~\cite{zhou2019twentyq} treat the case of a point-to-point channel. In~\cite{yavas2018random}, the authors derived achievable rates for random access channel coding using finite blocklength information theoretic tools~\cite{polyanskiy2010finite,TanBook}, where they considered a multiple access channel with unknown number of transmitters. Different from \cite{yavas2018random}, in our problem of searching for multiple targets, the corresponding channel coding problem is a virtual binary OR MAC, where the inputs are noiseless answers to queries and the noisy outputs are corrupted by query-dependent noise. Lemmas 1 and 2 of \cite{yavas2018random} facilitate the last step of our proof, but the bulk of the proof is new.

We remark that the multiple target search problem can be related to group testing, where the goal is to accurately recover a set of defective items out of a fixed number of items, using as few tests as possible. Dorfman formulated the problem of group testing~\cite{dorfman1943detection} and showed that pooling items together could significantly reduce the number of tests needed to accurately find the set of defective items. Our problem is related to the non-adaptive probabilistic group testing problem in the very sparse regime, where the number of defective items is finite, each item is assumed defective with a certain probability and each test is designed independently of other tests, when the query-dependent noise model reduces to a query-independent noise model and when each target random variable takes values in a one-dimensional discrete set. The information theoretic studies of the group testing problem~\cite{atia2012boolean,scarlett2018noisy,aldridge2018individual} establish upper and lower bounds on the number of tests needed to recover the defective items for the noiseless model and the noisy model with test-independent noise. In particular, strong converse theorems are established for various settings of group testing in the finite blocklength regime~\cite{johnson2017sc}. These results often are restricted to the sparse and linear regimes where the number of defective items scales with the total number of items (cf. \cite{scarlett2019group} for a recent survey). The theory of group testing cannot be directly applied to our problem, but the connection between the problems is interesting and worthwhile future work.

While this paper treats the case of non-adaptive queries, some of the results in this paper may be useful for the case of adaptive queries as well. For the case of a single target, in our previous work~\cite{zhou2019twentyq} we analyzed the second-order asymptotic performance of optimal non-adaptive query procedures and established a second-order asymptotic bound for an adaptive query procedure based on variable-length channel coding with feedback~\cite[Def. 1]{polyanskiy2011feedback}. Other adaptive query procedures proposed in the literature include: the three-stage random coding based algorithm~\cite{kaspi2018searching}, the sorted posterior matching (PM) algorithm~\cite{chiu2016sequential}, dyadic PM and hierarchical PM algorithms~\cite{chiu2021}. However, the second-order asymptotic performance of an optimal adaptive query procedure under the query-dependent noise model is unknown since a converse proof is challenging, as pointed out in~\cite{kaspi2018searching}. In a recent paper~\cite{zhou2021adaptive}, we compared the performance of the adaptive algorithm in \cite{zhou2019twentyq} and the sorted PM in~\cite{chiu2016sequential}, theoretically and numerically, with the conclusion that either algorithm could outperform the other one, depending on the query-dependent noise models. For the case of multiple targets, the only adaptive query algorithm we know of is a three-stage algorithm~\cite{kaspi2014multiple}, where each stage adjusts the search region according to previous stage(s) and then constructs a non-adaptive algorithm based on random coding. It would be interesting future work to derive similar second-order limits as obtained in this paper, which could result in low complexity practical algorithms for adaptive search for multiple targets.  An achievability result could be derived similarly to \cite{zhou2019twentyq} by borrowing ideas from variable length channel coding with feedback for a multiple access channel~\cite{rahul2016,kasper2014} but the converse part remains open. Low complexity algorithms might be constructed by generalizing the algorithms in \cite{chiu2016sequential,chiu2021} for a single target.

\section{Problem Formulation}
\subsection*{Notation}
Random variables and their realizations are denoted by upper case variables (e.g.,  $X$) and lower case variables (e.g.,  $x$), respectively. All sets are denoted in calligraphic font (e.g.,  $\mathcal{X}$).  We use $\Phi^{-1}(\cdot)$ to denote the inverse of the cumulative distribution function (cdf) of the standard Gaussian. We use $\bbR$, $\bbR_+$ and $\bbN$ to denote the sets of real numbers, positive real numbers and integers respectively. Given any two integers $(m,n)\in\bbN^2$, we use $[m:n]$ to denote the set of integers $\{m,m+1,\ldots,n\}$ and use $[m]$ to denote $[1:m]$. Given any $m\in\bbN$, for any length-$m$ vector $\ba=(a_1,\ldots,a_m)$, the infinity norm is defined as $\|\ba\|_{\infty}:=\max_{i\in[m]}|a_i|$ and the $L_2$ norm is defined as $\|\ba\|_2:=\sqrt{\sum_{i\in[m]}a_i^2}$. The set of all probability distributions on a finite set $\calX$ is denoted as $\calP(\calX)$ and the set of all conditional probability distributions from $\calX$ to $\calY$ is denoted as $\calP(\calY|\calX)$. Furthermore, we use $\calF(\calS)$ to denote the set of all probability density functions on the set $\calS$. Given $k\in\bbN$, random variables $(X_1,\ldots,X_k)\in\calX^k$ and a set $\calJ\subseteq[k]$, we use $X_\calJ$ to denote the collection of random variables $\{X_j\}_{j\in\calJ}$. Similarly, given $k$ length-$n$ random vectors $(X^n(1),\ldots,X^n(k))$, we use $X^n_{\calJ}$ to denote $\{X^n(j)\}_{j\in\calJ}$. We denote $x_\calJ$ and $x^n_{\calJ}$ as particular realizations of $X_\calJ$ and $X_\calJ^n$. All logarithms are base $e$ unless otherwise noted. Finally, we use $\bbo()$ to denote the indicator function.

\subsection{Problem Formulation under the Noisy 20 Questions Framework}
Given finite integers $(k,d)\in\bbN^2$, let $\bS^k:=(\bS_1,\ldots,\bS_k)$ be a sequence of $k$ independent $d$-dimensional random vectors, where for each $i\in[k]$, the random vector $\bS_i=(S_{i,1},\ldots,S_{i,d})$ is generated from an arbitrary probability density function (pdf) $f_{\bS}$ defined on the unit cube $[0,1]^d$. In this paper, we consider simultaneous search for multiple targets over the unit cube. Our results can be generalized to a search problem over any bounded $d$-dimensional region $\prod_{j\in[d]}[a_j,b_j]$, where $a_j$ and $b_j$ are finite real numbers.

In the problem of multiple target search, a player aims to accurately estimate the random location vectors $\bS^k$ of all $k$ targets by posing a sequence of queries with search regions $\calA^n=(\calA_1,\ldots,\calA_n)\subseteq([0,1]^d)^n$ to an oracle knowing $\bS^k$. After receiving the queries, the oracle generates binary yes/no answers $Z^n=(Z_1,\ldots,Z_n)$ that specify whether any target lies in search region $\calA_l$ for each $l\in[n]$. Given any sequence of queries $\calA^n\subseteq([0,1]^d)^n$ and $\bS^k$, for each $i\in[k]$, we denote $X^n(i)=(X_1(i),\ldots,X_n(i)):=(\bbo\{\bS_i\in\calA_1\},\ldots,\bbo\{\bS_i\in\calA_n\})$ as the $n$-dimensional binary answers to queries that ask whether the target $\bS_i$ lies in each query region. It follows that the noiseless answers $Z^n$ generated by the oracle is binary OR of $(X^n(1),\ldots,X^n(k))$, i.e., $Z^n=\{\bbo(\exists~i\in[k]:~X_1(i)=1),\ldots,\bbo(\exists~i\in[k]:~X_n(i)=1)\}$. Subsequently, these noiseless answers $Z^n$ are corrupted by noise via a binary input point-to-point query-dependent channel with transition matrix $P_{Y^n|Z^n}^{\calA^n}\in\calP(\calY^n|\{0,1\}^n)$, yielding noisy responses $Y^n=(Y_1,\ldots,Y_n)$. Given noisy responses $Y^n$, the player uses a decoding function to produce  estimates of the target location vectors $\bS^k$. Similarly to previous work~\cite{zhou2019twentyq,kaspi2018searching,chung2018unequal}, we assume that the alphabet $\calY$ for the noisy response is finite. Our problem formulation in this subsection largely parallels our treatment of the single target case in~\cite{zhou2019twentyq}.

\subsection{Description of The Query-Dependent Channel}
\label{sec:mdc}

We briefly describe the query-dependent channel that was partially discussed in~\cite{kaspi2018searching,chiu2016sequential,zhou2019twentyq}, which is also known as a channel with state~\cite[Chapter 7]{el2011network}. Given a sequence of queries $\calA^n\subseteq([0,1]^d)^n$, the channel from the oracle to the player is a memoryless channel whose transition probabilities are functions of the queries. Specifically, for any $(z^n,y^n)\in\{0,1\}^n\times\calY^n$,
\begin{align}
P_{Y^n|Z^n}^{\calA^n}(y^n|z^n)
&=\prod_{l\in[n]}P_{Y|Z}^{f(|\calA_l|)}(y_l|z_l)\label{def:mdchannel},
\end{align}
where $P_{Y|Z}^{f(|\calA_l|)}$ denotes the transition probability of a binary-input query-dependent noisy channel with finite output alphabet $\calY$, which depends on the query $\calA_l$ only through a Lipschitz continuous function $f(\cdot)$ of its size $|\calA_l|$. Specifically, given any query $\calA\subseteq[0,1]^d$, the size $|\calA|$ of $\calA$ is defined as its Lebesgue measure, i.e., $|\calA|=\int_{t\in\calA}\rmd t$. The function $f:[0,1]\to\bbR_+$ is a bounded Lipschitz continuous function with parameter $\mu$, i.e., $|f(p_1)-f(p_2)|\leq \mu |p_1-p_2|$ for any $(p_1,p_2)\in[0,1]^2$ and $\max_{q\in[0,1]}f(q)<\infty$. A simple example of $f(\cdot)$ is the identity function, i.e., $f(|\calA|)=|\calA|$ and for this case $\mu=1$. Furthermore, if $f(\cdot)$ is a constant value function, then the channel reduces to a query-independent channel. Thus, the channel model above unifies both query-independent and query-dependent noisy channels.

Let $q\in[0,1]$ be arbitrary. Consider any $\xi\in(0,\min(q,1-q))$. Similarly to \cite{zhou2019twentyq}, we assume that the noisy channel is continuous in the sense that there exists a positive constant $c(q)$ depending on $q$ only such that
\begin{align}
\max\Bigg\{&\bigg\|\bigg\{\log\frac{P_{Y|Z}^q(y,z)}{P_{Y|Z}^{{q+\xi}}(y,z)}\bigg\}_{(y,z)\in\calY\times\calZ}\bigg\|_{\infty},\bigg\|\bigg\{\log\frac{P_{Y|Z}^q(y,z)}{P_{Y|Z}^{{q-\xi}}(y,z)}\bigg\}_{(y,z)\in\calY\times\calZ}\bigg\|_{\infty}\Bigg\}\leq c(q)\xi\label{assump:continuouschannel}.
\end{align}

Examples of the query-dependent noisy channels are given as follows\footnote{This is an generalization of the channels for the special case of $f(|\calA|)=|\calA|$ that appeared in~\cite{zhou2019twentyq,kaspi2014multiple}.}.
\begin{definition}
\label{def:mdBSC}
Given any $\calA\subseteq[0,1]$, a channel $P_{Y|Z}^{f(|\calA|)}$ is said to be a query-dependent binary symmetric channel (BSC) if $\calZ=\calY=\{0,1\}$ and $\forall~(y,z)\in\{0,1\}^2$,
\begin{align}
P_{Y|Z}^{f(|\calA|)}(y|z)=(f(|\calA|))^{\bbo(y\neq z)}(1-f(|\calA|))^{\bbo(y=z)}.
\end{align}
\end{definition}
This definition generalizes \cite[Theorem \ref{theorem:fbl:achievability}]{kaspi2018searching}, where the authors considered a query-dependent BSC with the function $f(|\calA|)=|\calA|$. For a query-dependent BSC, the output bit is the same as the input with probability $1-f(|\calA|)$ and the output bit flips the input with probability $f(|\calA|)$. It can be verified that the constraint in \eqref{assump:continuouschannel} is satisfied for the query-dependent BSC. For this case, a valid choice of the function $f(\cdot)$ should satisfy $f(|\calA|)\leq 1$ for any $\calA\subseteq[0,1]$. In particular, we only consider Lipschitz continuous functions $f(\cdot)$ such that $f(|\calA|)\leq \frac{1}{2}$ for any $\calA\subseteq[0,1]$, ensuring that the crossover probability of the BSC is not greater than $\frac{1}{2}$.

\begin{definition}
\label{def:mdBEC}
Given any $\calA\subseteq[0,1]$, a query-dependent channel $P_{Y|Z}^{f(|\calA|)}$ is said to be a query-dependent binary erasure channel (BEC) if $\calZ=\{0,1\}$, $\calY=\{0,1,\rme\}$ and for any $(y,z)\in\calY\times\calZ$,
\begin{align}
P_{Y|Z}^{f(|\calA|)}(y|z)=(1-f(|\calA|))^{\bbo(y=z)}(f(|\calA|))^{\bbo(y=\rme)}
\end{align}
\end{definition}
The output bit of a query-dependent BEC is erased with probability $f(|\calA|)$ and thus the Lipschitz continuous function $f(\cdot)$ is chosen such that $f(|\calA|)<1$.

Each of these query-dependent channels will be used to illustrate our results. In subsequent subsections, these channel characterizations of the multiple target search problem will be related to equivalent multiple access channels, specifically the OR MAC channel where the input $Z$ is the binary OR of multiple binary inputs.

\subsection{Definition of the Fundamental Limit}

\label{sec:search:ktarget}

A non-adaptive query procedure for multiple target search is defined as follows.
\begin{definition}
\label{def:procedure:ktarget}
Given any $(n,k,d)\in\bbN^3$, $\delta\in\bbR_+$ and $\varepsilon\in[0,1]$, a $(n,k,d,\delta,\varepsilon)$-non-adaptive query procedure for noisy 20 questions consists of 
\begin{itemize}
\item $n$ queries $(\calA_1,\ldots,\calA_n)$ where for each $l\in[n]$, $\calA_l\subseteq[0,1]^d$,
\item and a decoder $g:\calY^n\to\calS_m\subseteq ([0,1]^d)^m,~m\leq k$,
\end{itemize}
such that the excess-resolution probability satisfies
\begin{align}
\rmP_\rme(n,k,d,\delta)
&:=\sup_{f_{\bS}\in\calF([0,1]^d)}\max\bigg\{\Pr\Big\{\exists~i\in[k]:~\min_{\hat{\bs}\in\calS_m}|\hat{\bs}-\bS_i|_{\infty}>\delta\Big\},\Pr\Big\{\exists~\hat{\bs}\in\calS_m:\min_{i\in[k]}|\hat{\bs}-\bS_i|_{\infty}>\delta\Big\}\bigg\}\leq \varepsilon\label{def:excessresolution3},
\end{align}
where $m$ is the estimated number of targets and $\calS_m=\{\hat{\bs}_1,\ldots,\hat{\bs}_m\}$ is the set of estimated location vectors output by the decoder.
\end{definition}
Note that the estimated number of targets $m$ could be smaller than the actual number of targets $k$ when two or more targets are close to each other with respect to the target resolution $\delta$. When $\varepsilon$ is small, the first probability term on the right side of the equality in \eqref{def:excessresolution3} ensures that the set of reproduced target locations $\calS_m$ is \emph{sufficient} in the sense that for every target location vector $\bS_i$, we can find a vector $\hat{\bs}$ in the estimated set $\calS_m$ such that the estimate is accurate within resolution $\delta$; the second probability term ensures that $\calS_m$ is \emph{non-redundant} in the sense that every vector $\hat{\bs}$ in $\calS_m$ lies in the proximity of at least one target location vector $\bS_i$. 

The minimal achievable resolution of an optimal non-adaptive query procedure is then defined as
\begin{align}
\delta^*(n,k,d,\varepsilon)&:=\inf\{\delta:\exists~\mathrm{an~}(n,k,d,\delta,\varepsilon)\mathrm{-non}\mathrm{-adaptive}\mathrm{~query}\mathrm{~procedure}\}\label{def:delta*}.
\end{align}
We remark that the minimal achievable sample complexity $n^*(\delta,k,d,\varepsilon)$ can be defined analogously to \eqref{def:delta*}, and it can be expressed as a function of $\delta^*(n,k,d,\varepsilon)$. Thus, we focus on $\delta^*(n,k,d,\varepsilon)$ in this paper.

\section{Main Results and Discussions}

\subsection{Preliminary Definitions}
To present our results and proposed query procedures, the following definitions are needed. Given any $X_{[k]}=(X_1,\ldots,X_k)$, we use the symbol $Z:=\{\exists~i\in[k]:~X_i=1\}$ to denote the binary OR of $X_{[k]}$. For a realization $x_{[k]}$ of $X_{[k]}$, $z$ is a realization of $Z$. Given any $p\in[0,1]$, for any $(x_{[k]},y)\in[0,1]^k\times\calY$, define the joint distribution
\begin{align}
P_{X_{[k]}Y}^{f(p),k}(x_{[k]},y)
&:=\Big(\prod_{i\in[k]}P_X(x_i)\Big)P_{Y|Z}^{f(p)}(y|z)\label{def:pjointk},
\end{align}
where $P_X$ denotes the Bernoulli distribution with parameter $p$ and $P_{Y|Z}^{f(p)}$ denotes the binary input point-to-point query-dependent channel. We establish in Appendix \ref{just} that the induced noisy channel (conditional distribution) $P_{Y|X_{[k]}}^{f(p),k}$ corresponds to a binary input multiple access channel  that satisfies the permutation-invariant, reducibility, friendliness and interference assumptions in \cite{yavas2018random}. We remark that the MAC $P_{Y|X_{[k]}}^{f(p),k}$ is degenerate since it is a point-to-point channel with the input $Z$ being the binary OR of inputs $X_{[k]}=(X_1,\ldots,X_k)$.

Given any $t\in[k]$, for any $\calJ\subset[t]$ and any $p\in[0,1]$, define the (conditional) mutual information density
\begin{align}
\imath_\calJ^{f(p),t}(x_{[t]};y)
&:=\log\frac{P_{Y|X_{[t]}}^{f(p),t}(y|x_{[t]})}{P_{Y|X_{\calJ}}^{f(p),t}(y|x_\calJ)}\label{def:cdmi},
\end{align}
where the distributions $P_{Y|X_{[t]}}^{f(p),t}$ and $P_{Y|X_{\calJ}}^{f(p),t}$ are both induced by a joint distribution $P_{X_{[t]}Y}^{f(p),t}$, which is defined similarly to $P_{X_{[k]}Y}^{f(p),k}$ except that $k$ is replaced by $t$. Note that when $\calJ=\emptyset$, $\imath_\emptyset^{f(p),t}(x_{[t]};y)$ is the mutual information density, i.e.,
\begin{align}
\imath_\emptyset^{f(p),t}(x_{[t]};y)
&:=\log\frac{P_{Y|X_{[t]}}^{f(p),t}(y|x_{[t]})}{P_{Y}^{f(p),t}(y)}\label{def:cdmiempty}.
\end{align}
The mutual information density $\imath_\calJ^{f(p),t}(x_{[t]};y)$ is critical in the presentation and proof of our results. In particular, the mean and the variance of $\imath_\calJ^{f(p),t}(\cdot)$ prove and parametrize our second-order asymptotic result in Theorem \ref{result:second:ktarget}:
\begin{align}
C_\calJ(p,t)&:=\mathbb{E}_{P_{X_{[t]}Y}^{f(p),t}}[\imath_\calJ^{f(p),t}(X_{[t]};Y)]=I(X_{[t]\setminus\calJ};Y|X_\calJ)\label{def:cjpt},\\
V_\calJ(p,t)&:=\mathrm{Var}_{P_{X_{[t]}Y}^{f(p),t}}[\imath_\calJ^{f(p),t}(X_{[t]};Y)]\label{def:vjpt}.
\end{align}
When $\calJ=\emptyset$, $C_\emptyset(p,t)$ and $\rmV_\emptyset(p,t)$ are the mutual information and the variance of $\imath_\emptyset^{f(p),t}(X_{[t]};Y)$ respectively, i.e., 
\begin{align}
C_\emptyset(p,t)&=I(X_{[t]};Y)\label{def:cemptypt},\\
\rmV_\emptyset(p,t)&=\mathrm{Var}[\imath_\emptyset^{f(p),t}(X_{[t]};Y)]\label{def:vemptypt}.
\end{align}
Define the capacity of the binary-input MAC induced by \eqref{def:pjointk} as
\begin{align}
C(k)
&=:\max_{p\in(0,1)}C_\emptyset(p,k)\label{def:maxpc},
\end{align}
where we use $C_\emptyset(p,k)$ to denote the mutual information $I(X_{[k]};Y)$ and the joint distribution of $(X_{[k]};Y)$ was given in \eqref{def:pjointk}. Let $\calP_k$ denote the set of capacity achieving optimizers $p^*$ in \eqref{def:maxpc}. The dispersion for the channel is then defined as
\begin{align}
\rmV(k,\varepsilon)
&:=\left\{
\begin{array}{cc}
\min_{p^*\in\calP_k}\rmV_\emptyset(p,k)&\mathrm{if~}\varepsilon\leq 0.5,\\
\max_{p^*\in\calP_k}\rmV_\emptyset(p,k)&\mathrm{if~}\varepsilon>0.5.
\end{array}
\right.\label{def:dispersion}
\end{align}
As we show in Theorem \ref{result:second:ktarget}, $C(k)$ and $\rmV(k,\varepsilon)$ characterize our second-order asymptotic result. Since we consider finite input and output alphabets, similar to \cite[Lemma 47]{polyanskiy2010finite}, the dispersion $\rmV(k,\varepsilon)$ is finite.

For any $\gamma>0$, given each $t\in[k]$ and $\calJ\subset[t]$, define the following set of sequences
\begin{align}
\calD_\calJ^{n,t}(\gamma)
:=\bigg\{(x^n_{[t]},y^n)\in(\{0,1\}^n)^t\times\calY^n:\sum_{l\in[n]}\imath_\calJ^{f(p),t}(x_{l,[t]};y_l)>d(t-|\calJ|)\log M+\gamma\bigg\},\label{def:calDj}
\end{align}
and define
\begin{align}
\calD^{n,t}(\gamma)&:=\bigcap_{\calJ\subset[t]}\calD_\calJ^{n,t}(\gamma)\label{def:calD},
\end{align}
where for each $l\in[n]$, $x_{l,[t]}$ is the collection of $l$-th elements of all $t$ sequences $x^n_{[t]}=(x^n(1),\ldots,x^n(t))$, i.e., $x_{l,[t]}=(x_{l,1},\ldots,x_{l,t})$. The sets $\calD^{n,t}(\gamma)$ collect encoded noiseless and noisy response pairs $(x_{[t]}^n,y_n)$ whose average mutual information densities over $n$ queries exceeds $\gamma$-dependent thresholds when the number of unique quantized indices is $t$. These sets are used by the decoder of our non-adaptive query procedure in Algorithm \ref{procedure:nonadapt:ktaget} and they play a crucial role in establishing the non-asymptotic achievability bound in Theorem \ref{theorem:fbl:achievability}. Such sets arise in the information spectrum method for MAC analysis , e.g., \cite[Lemma 7.10.1]{han2003information}. We derive an achievability result by relating the current problem of search for multiple targets to data transmission over a MAC. The sets $\calD^{n,t}(\gamma)$ are thus critical. If $x_{[t]}^n=(x_1^n,\ldots,x_t^n)$ and $y^n$ are the transmitted and received codewords of a MAC, corresponding to our encoded noiseless and the noisy responses, the accumulated (conditional) mutual information densities $\sum_{l\in[n]}\imath_\calJ^{f(p),t}(x_{l,[t]};y_l)$ are large with high probability and the inequality in $\calD_\calJ^{n,t}(\gamma)$ is satisfied with properly chosen $M$ and $\gamma$.

Finally, for any $(k,M)\in\bbN^2$, let $\calL(k,M)$ be the set of all length-$k$ vectors whose elements are ordered increasingly and each element takes values in $[M]$, i.e.,
\begin{align}
\calL(k,M)
&:=\{(i_1,\ldots,i_k)\in[M]^k:~\forall~(j,l)\in[k]^2~,j<l\rightarrow i_j<i_l\}
\label{def:calL}.
\end{align}
The set $\calL(k,M)$ will be critical in the design and analysis of our non-adaptive query procedure.

\subsection{Non-Asymptotic Bounds}
In this section, we provide non-asymptotic bounds on the excess-resolution probability. Recall the definition of the joint distribution $P_{X_{[k]} Y}^{f(p),k}$ in \eqref{def:pjointk}.

Our first result characterizes the performance of the non-adaptive query procedure with finitely many queries. Recall that $f(\cdot)$ is a bounded Lipschitz continuous function with parameter $\mu$.
\begin{theorem}
\label{theorem:fbl:achievability}
Given finite integers $(n,k,d)\in\bbN^3$ and any query-dependent channel satisfying \eqref{assump:continuouschannel}, for any $(M,p,\eta,\gamma)\in\bbN\times(0,1)\times\bbR_+^2$, there exists a $(n,k,d,\frac{1}{M},\varepsilon)$-non-adaptive query procedure where
\begin{align}
\varepsilon
\leq 4n\exp(-2M^d\eta^2)+\exp(n\mu\eta c(f(p)))\bigg((k+1)2^k\exp(-\gamma)+\max_{t\in[k]}\Pr_{(P_{X_{[t]} Y}^{f(p),t})^n}\bigg\{(X^n_{[t]},Y^n)\notin\bigcup_{\calJ\subset[t]}\calD_{\calJ}^{n,t}(\gamma)\bigg\}\bigg)\label{ach:fbl:eqn}.
\end{align}
\end{theorem}

The proof of Theorem \ref{theorem:fbl:achievability} is provided in Section IV where we combine a random coding argument, the information spectrum method for the multiple access channel~\cite{han2003information} and the change-of-measure technique~\cite{csiszar2011information,polyanskiy2011feedback}. Specifically, we use the non-adaptive query procedure in Algorithm \ref{procedure:nonadapt:ktaget} below. In this algorithm, we partition the unit cube into $M^d$ disjoint sub-cubes $\{\calS_1,\ldots,\calS_{M^d}\}$ and generate random vectors $(x^n(1),\ldots,x^n(M^d))$. For each $l\in[n]$, the $l$-th query asks whether any targets lie in the search region $\calA_l:=\bigcup_{i\in[M^d]:x_l(i)=1}\calS_i$. In the decoding part, the information density threshold rules in \eqref{def:calD} are used. To analyze the performance of Algorithm \ref{procedure:nonadapt:ktaget}, for any query vectors $(x^n(1),\ldots,x^n(M^d))$, we first identify three error events that lead to excess-resolution. Subsequently, similar to the achievability proof for channel coding problems~\cite{cover2012elements}, we adopt the random coding argument by assuming that the query vectors are random and then we bound the ensemble excess-resolution probability, which is also averaged over the randomness of query vectors. Then, the ensemble error probability is related to the error probability of data transmission over a MAC and the information spectrum method~\cite{han2003information} is applied to yield the desired bound in Theorem \ref{theorem:fbl:achievability}. Finally, the existence of deterministic query vectors result from the fact that for any random variable $Z$ and real number $a\in\bbR$, $\mathbb{E}[Z]<a$ implies that there exists a realization $z$ of the random variable $Z$ such that $z<a$.

We next describe what each term in Theorem \ref{theorem:fbl:achievability} represents. The first term $4n\exp(-2M^d\eta)$ results from the atypicality of the query-dependent channel. In particular, if the query-dependent noisy channel is too different from the target channel $(P_{Y|X_{[t]}}^{f(p),t})^n$, which is induced by the joint distribution $P_{X_{[t]}Y}^{f(p),t}$, an error is declared in our analysis and this probability is upper bounded by $4n\exp(-2M^d\eta^2)$. The multiplicative term $\exp(n\mu\eta c(f(p)))$ results from the change-of-measure technique, the assumption \eqref{assump:continuouschannel} on the channel and the assumption that $f(\cdot)$ is Lipschitz continuous. Specifically, when the query-dependent noisy channel is close to $(P_{Y|X_{[t]}}^{f(p),t})^n$, we use a change-of-measure technique to calculate the probability of error events under $(P_{Y|X_{[t]}}^{f(p),t})^n$ instead of the true query-dependent noisy channel. The cost of this measure change is quantified by the multiplicative term $\exp(n\mu\eta c(f(p)))$.  Finally, the terms inside the parenthesis of \eqref{ach:fbl:eqn} result from the error probability analysis of data transmission over the memoryless MAC $(P_{Y|X_{[t]}}^{f(p),t})^n$. The term $(k+1)2^k\exp(-\gamma)$ upper bounds the probability that incorrect messages are decoded and the remaining term upper bounds the probability that the transmitted messages do not satisfy the decoding criterion (refer to the while loop in Algorithm \ref{procedure:nonadapt:ktaget}). In particular, the maximum over $t\in[k]$ is taken to control the worst case excess-resolution probability. This is because each query in Algorithm \ref{procedure:nonadapt:ktaget} asks whether there are any targets in the union of disjoint sub-cubes. The index of the sub-cube where each target locates is analogous to a message sent by a transmitter in the MAC. If two targets are close to each other, they fall into the same sub-cube and the total number of messages is less than $k$. However, different from coding problems over a MAC, in 20 questions estimation for multiple targets, we only require the recovery of the set of target locations according to \eqref{def:excessresolution3}, where all messages are sent simultaneously. This corresponds to an unordered recovery of the messages, whose number is random and depends on the target location vectors $\bS^k=(\bS_1,\ldots,\bS_k)\in([0,1]^d)^k$.

\begin{algorithm}[bt]
\caption{Non-adaptive query procedure for multiple target search}
\label{procedure:nonadapt:ktaget}
\begin{algorithmic}
\REQUIRE The number of queries $n\in\bbN$ and three parameters $(M,p,\gamma)\in\bbN\times(0,1)\times\bbR_+$
\ENSURE A set of estimated target location vectors $\calS_m\subseteq([0,1]^d)^k$.

\hrulefill
\STATE Partition the unit cube of dimension $d$ into $M^d$ equal-sized disjoint cubes $(\calS_1,\ldots,\calS_{M^d})$.
\STATE \emph{Query generation}: 
\STATE Generate $M^d$ binary vectors $(x^n(1),\ldots,x^n(M^d))$, where for each $i\in[M^d]$ and $l\in[n]$, $x_l(i)$ is generated independently from a Bernoulli distribution with parameter $p$.
\STATE Form query regions $\calA_l$ for each $l\in[n]$ as
\begin{align*}
\calA_l:=\bigcup_{i\in[M^d]:x_l(i)=1}\calS_i
\end{align*}
\STATE \emph{Answer collection:}
\STATE $l \leftarrow 1$.
\WHILE{$l\leq n$}
\STATE Submit the $l$-th query that asks the oracle whether there are targets in the region $\calA_l$.
\STATE Obtain a noisy response $y_l$ from the oracle.
\STATE $l \leftarrow l+1$.
\ENDWHILE
\STATE \emph{Decoding:}
\STATE Initialize $\calS_m$ as the empty set.
\STATE $t \leftarrow k$.
\WHILE{$t>0$}
\IF{$\exists$ a tuple $(i_1,\ldots,i_t)\in\calL(t,M^d)$ such that $(x^n(i_1),\ldots,x^n(i_t),y^n)\in\calD^{n,t}(\gamma)$ (cf. \eqref{def:calD})}
\STATE Return $\calS_m$ as the centers of cubes $\calS_{i_j}$ with $j\in[t]$
\STATE $t\leftarrow 0$
\ELSE
\STATE $t \leftarrow t-1$.
\ENDIF
\ENDWHILE
\end{algorithmic}
\end{algorithm}

We next present a converse result, which lower bounds the resolution of any non-adaptive query procedure subject to a constraint on the excess-resolution probability. 
\begin{theorem}
\label{theorem:fbl:converse}
Given any $(n,k,d,\varepsilon)\in\bbN^3\times(0,1)$, for any $\beta\in(0,\frac{\varepsilon}{2}]$ and $\kappa\in(0,1-\varepsilon-2k^2d\beta)$, the minimal achievable resolution $\delta^*(n,k,d,\varepsilon)$ satisfies 
\begin{align}-\log \delta^*(n,k,d,\varepsilon)
\leq \sup_{\calA^n\in([0,1]^d)^n}\frac{\sup\bigg\{\psi\Big|\Pr\Big\{\sum_{l\in[n]}\imath_{\calA_l}(Z_l;Y)\leq \psi\Big\}\leq \varepsilon+2k^2d\beta+\kappa\bigg\}-\log\kappa-dk\log\beta}{dk}\label{fbl:conversebd}.
\end{align}
where $Z_l$ denotes the Bernoulli random variable with parameter $1-(1-|\calA_l|)^k$, $P_{Y_l}^{\calA_l}$ is the induced marginal distribution on the alphabet $\calY$ by $P_{Z_l}$ and the query-dependent noisy channel $P_{Y|Z}^{f(|\calA_l|)}$ and $\imath_{\calA_l}(z;y)$ is a query-dependent mutual information density defined as follows:
\begin{align}
\imath_{\calA_l}(z;y):=\log\frac{P_{Y|Z}^{f(|\calA_l|)}(y|z)}{P_{Y_l}^{\calA_l,t}(y)}\label{def:mutualyz}.
\end{align}
\end{theorem}

The proof of Theorem \ref{theorem:fbl:converse} is provided in Section \ref{proof:fbl:con}. To prove Theorem \ref{theorem:fbl:converse}, we first lower bound the excess-resolution probability of a non-adaptive query procedure with the error probability of estimating quantized indices of the target location vectors that are independently and uniformly distributed over the unit cube. Subsequently, we show that the latter problem is closely related to data transmission over a point-to-point channel and adapt the finite blocklength converse bound in~\cite[Proposition 4.4]{TanBook}, similarly to \cite[Theorem 2]{zhou2019twentyq}.

Theorem \ref{theorem:fbl:converse} characterizes the best achievable resolution of any non-adaptive query procedure using $n$ queries that searches for $k$ targets simultaneously. Analogously to Theorem \ref{theorem:fbl:achievability}, the dominant probability term in Theorem \ref{theorem:fbl:converse} results from the converse proof of the reliability of data transmission over a noisy channel and the maximum over $t\in[k]$ accounts for each possible number of quantized indices. The parameter $\beta$ is related to the definition of the quantization function that is used to quantize the continuous value target location vector and $\gamma$ is a parameter to be optimized. Since an optimization over all possible queries $\calA^n\in([0,1]^d)^n$ is involved, the non-asymptotic bound in Theorem \ref{theorem:fbl:converse} is difficult to calculate. However, as demonstrated in the proof of Theorem \ref{result:second:ktarget}, for $n$ sufficiently large, the bound in Theorem \ref{theorem:fbl:converse} can be approximated with simple equations independent of the queries $\calA^n$, and such approximations are collectively known as second-order asymptotics~\cite{TanBook}.

\subsection{Second-Order Asymptotic Approximation and Further Discussions}

Recall the definitions of $C(k)$ in \eqref{def:maxpc} and $\rmV(k,\varepsilon)$ in \eqref{def:dispersion}.
\begin{theorem}
\label{result:second:ktarget}
For any finite numbers $(k,d)\in\bbN^2$ and any $\varepsilon\in(0,1)$, the achievable resolution $\delta^*(n,k,d,\varepsilon)$ of an optimal non-adaptive query procedure satisfies
\begin{align}
-\log \delta^*(n,k,d,\varepsilon)=\frac{nC(k)+\sqrt{n\rmV(k,\varepsilon)}\Phi^{-1}(\varepsilon)+\Theta(\log n)}{dk}\label{eqn:mainresult},
\end{align}
where the remainder term in \eqref{eqn:mainresult} satisfies
\begin{align}
-\frac{1}{2}\log n+O(1)\leq \Theta(\log n)\leq \frac{1}{2}\log n+O(1).
\end{align}

\end{theorem}
The proof of Theorem \ref{result:second:ktarget} is provided in Section \ref{proof:result:second}. In our achievability proof of Theorem \ref{result:second:ktarget}, we use the non-adaptive query procedure in Algorithm \ref{procedure:nonadapt:ktaget} and thus prove its second-order asymptotic optimality.

\begin{figure}[tb]
\centering
\begin{tabular}{cc}
\includegraphics[width=.45\columnwidth]{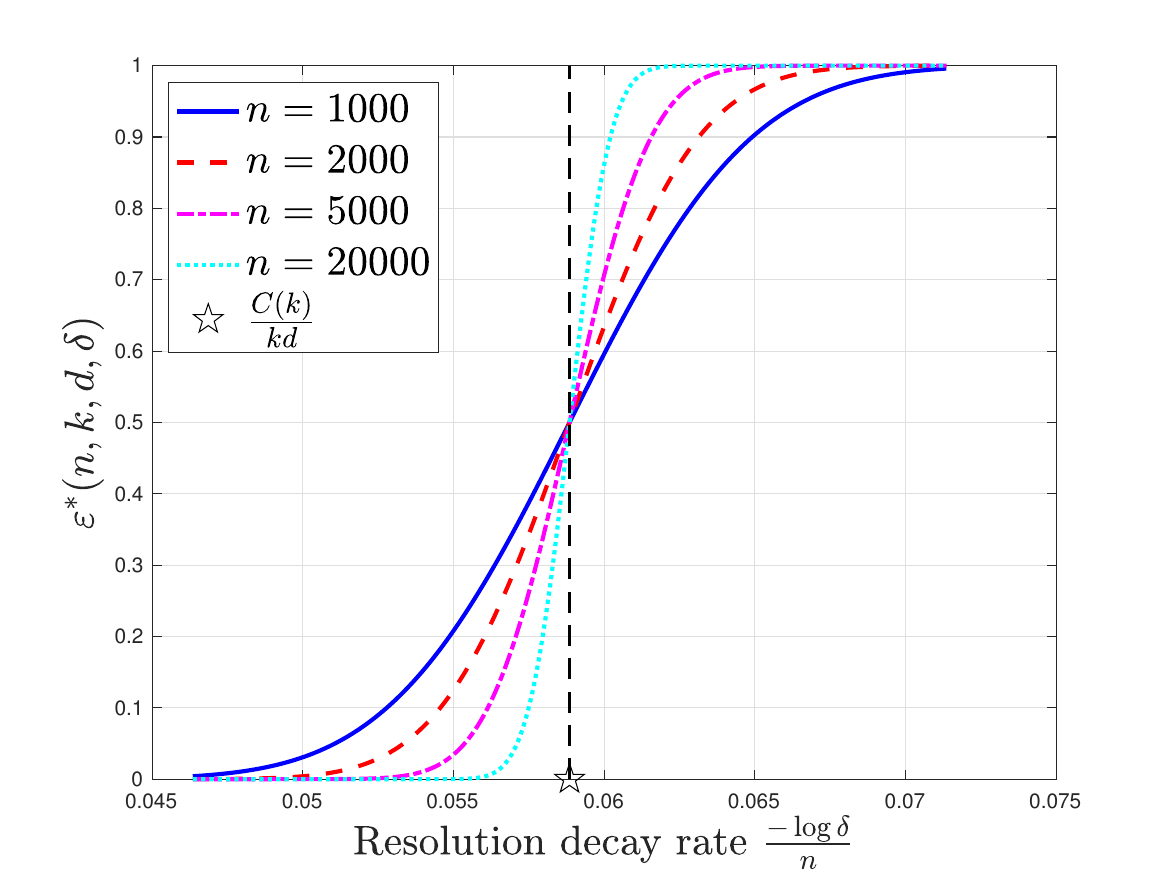}&\includegraphics[width=.45\columnwidth]{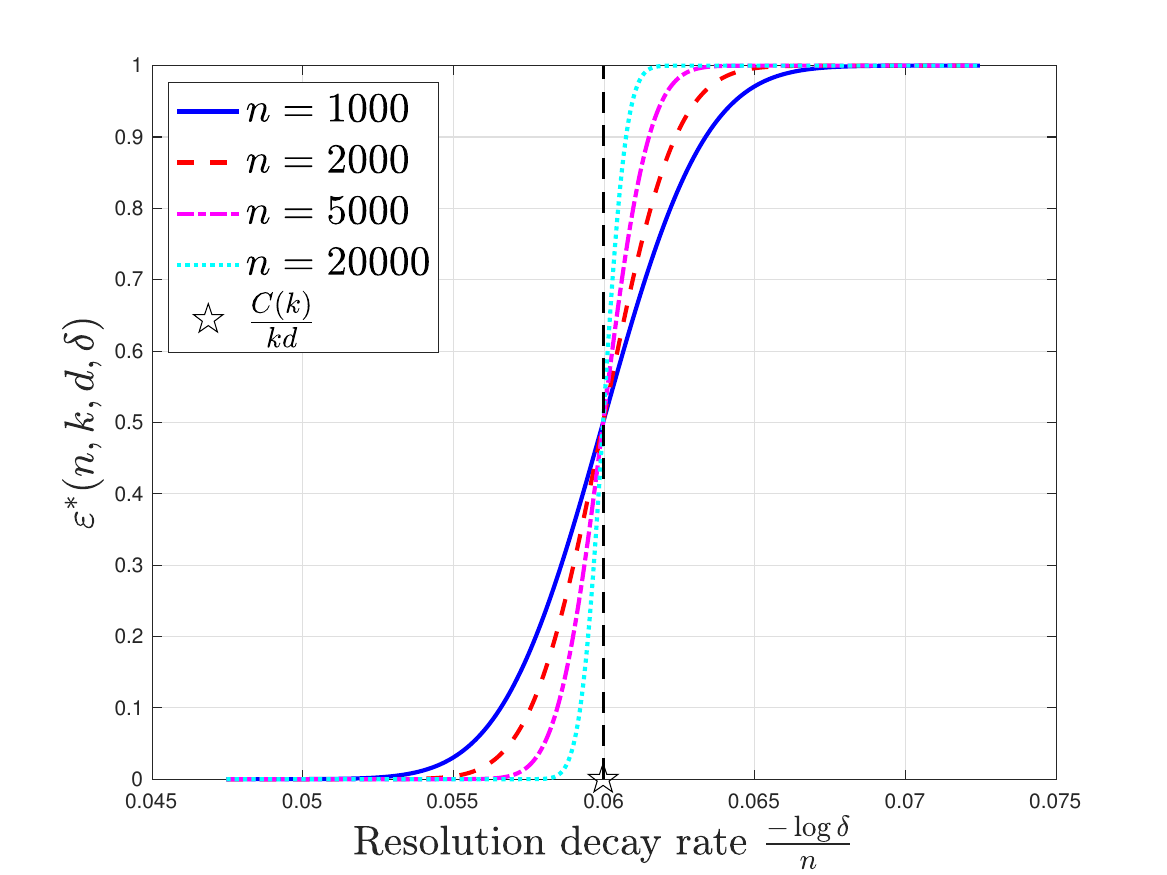}\\
{(a) BSC with $f(p)=0.3+0.1p$} & { (b) BEC with $f(p)=0.6+0.2p$}
\end{tabular}
\caption{Illustration of phase transition for an optimal non-adaptive query procedure that searches for $k=2$ targets over the unit cube of dimension $d=2$ for different Lipschitz continuous channel functions $f(\cdot)$. The pentagram denotes the critical threshold $\frac{C(k)}{dk}$.}
\label{illus_pt_ktarget}
\end{figure}

We make several remarks. If we let $\varepsilon^*(n,k,d,\delta)$ be the minimal excess-resolution probability of any non-adaptive query procedure, then Theorem \ref{result:second:ktarget} implies that for $n$ sufficiently large,
\begin{align}
\varepsilon^*(n,k,d,\delta)=\Phi\left(\frac{-dk\log\delta-nC(k)}{\sqrt{n\rmV(k,\varepsilon)}}\right)+o(1)\label{phaset}.
\end{align}
Note that \eqref{phaset} implies a phase transition phenomenon. Specifically, if the target resolution decay rate $\frac{-\log\delta}{n}$ is strictly greater than $\frac{C(k)}{dk}$, then the excess-resolution probability tends to \emph{one} as the number of queries $n$ tends to infinity; on the other hand, when the target resolution decay rate is strictly less than the critical rate $\frac{C(k)}{dk}$, then the excess-resolution probability \emph{vanishes} as the number of queries $n$ increases. This sharp asymptotic transition of the excess-resolution probability between zero and one as a function of the resolution decay rate is called a phase transition. See Figure \ref{illus_pt_ktarget} for an illustration.

From Theorem \ref{result:second:ktarget}, it follows that for any $\varepsilon\in(0,1)$, the resolution decay rate of an optimal non-adaptive query procedure satisfies
\begin{align}
 \lim_{n\to\infty}\frac{-\log\delta^*(n,k,d,\varepsilon)}{n}&=\frac{C(k)}{dk}\label{altexp}.
\end{align}
Eq. \eqref{altexp} is a strong converse result that refines the classical weak converse argument, which is only valid for $\varepsilon\to 0$. Note that the strong converse and phase transition complement each other. Specifically, the phase transition establishes that the asymptotic excess-resolution probability equals either $0$ or $1$, depending on whether the resolution decay rate is above or below the critical threshold $\frac{C(k)}{dk}$. In contrast, the strong converse asserts that the asymptotic resolution decay rate equals $\frac{C(k)}{dk}$ if the excess-resolution probability takes values in $(0,1)$.

We remark that an alternative characterization of $\frac{C(k)}{dk}$ in \eqref{altexp} is
\begin{align}
\frac{C(k)}{dk}=\max_{p\in(0,1)}\min_{t\in[k]}\frac{C_\emptyset(p,t)}{dt}\label{ctptmin}.
\end{align}
The relation in~\eqref{ctptmin} is critical in our proof of Theorem \ref{result:second:ktarget}. To obtain the result in~\eqref{ctptmin}, we find that the inequality in \cite[Lemma 1]{yavas2018random}, which was derived for random access channel coding, is helpful. Since the MAC channel induced by the joint distribution in \eqref{def:pjointk} satisfies the conditions in \cite[Lemma 1]{yavas2018random} (see Appendix \ref{just}), it follows that $\frac{C_\emptyset(p,k)}{k}<\frac{C_\emptyset(p,t)}{t}$ for any $t<k$ and thus $\min_{t\in[k]}\frac{C_\emptyset(p,t)}{dt}=\frac{C_\emptyset(p,k)}{dk}$.

We further discuss the role of \eqref{ctptmin} in the achievability proof. The minimization over $t\in[k]$ in \eqref{ctptmin} follows from the fact that given a certain resolution, the number of sub-cubes containing targets might be fewer than the total number of targets $k$. This is because in Algorithm \ref{procedure:nonadapt:ktaget} the unit cube is partitioned into equal-sized disjoint sub-cubes to quantize the target locations $\bs^k=(\bs_1,\ldots,\bs_k)$. Two targets $(\bs_i,\bs_j)$ might be quantized into the same sub-cube if they are closer to each other than the given resolution $\delta$. To ensure that our result holds for all possible cases, a minimization over the number of quantized targets accounts for the worst case. Furthermore, to maximize the performance of the non-adaptive query procedure, we choose the second-order asymptotically optimal codebook by maximizing over the parameter $p\in(0,1)$. Such intuition is confirmed to be second-order asymptotically optimal by our converse proof of Theorem \ref{result:second:ktarget}.

Finally, if one uses the $L_2$ norm instead of the $L_\infty$ norm as the performance criterion in the definition of the excess-resolution probability in \eqref{def:excessresolution3}, we can define the corresponding fundamental limit as $\delta^*_{L_2}(n,k,d,\varepsilon)$. We then have $\delta^*_\infty(n,k,d,\varepsilon)\leq \delta^*_{L_2}(n,k,d,\varepsilon)\leq \sqrt{d}\delta^*(n,k,d,\varepsilon)$ and thus the second-order asymptotic result in Theorem \ref{result:second:ktarget} also holds under the $L_2$ norm performance criterion when the dimension $d$ is finite. This is because given any target location vector $\bs=(s_1,\ldots,s_d)\in[0,1]^d$ and any estimated vector $\hat{\bs}=(\hats_1,\ldots,\hats_d)\in[0,1]^d$, we have $\|\hat{\bs}-\bs\|_\infty\leq \|\hat{\bs}-\bs\|_2\leq \sqrt{d}\|\hat{\bs}-\bs\|_\infty$. Readers can refer to \cite[Section II.C]{zhou2019twentyq} for discussions on the relationship between results for $L_2$ and $L_\infty$ norms for the special case of $k=1$.

\subsection{Specializations to Query-Dependent BSC and BEC}
\label{sec:special}
We specialize our second-order asymptotic results in Theorem \ref{result:second:ktarget} to query-dependent versions of the BSC and BEC by calculating the capacity $C(k)$ and the dispersion $\rmV(k,\varepsilon)$. 

\subsubsection{Case of a query-dependent BSC}
\begin{figure}[tb]
\centering
\begin{tabular}{cc}
\includegraphics[width=.45\columnwidth]{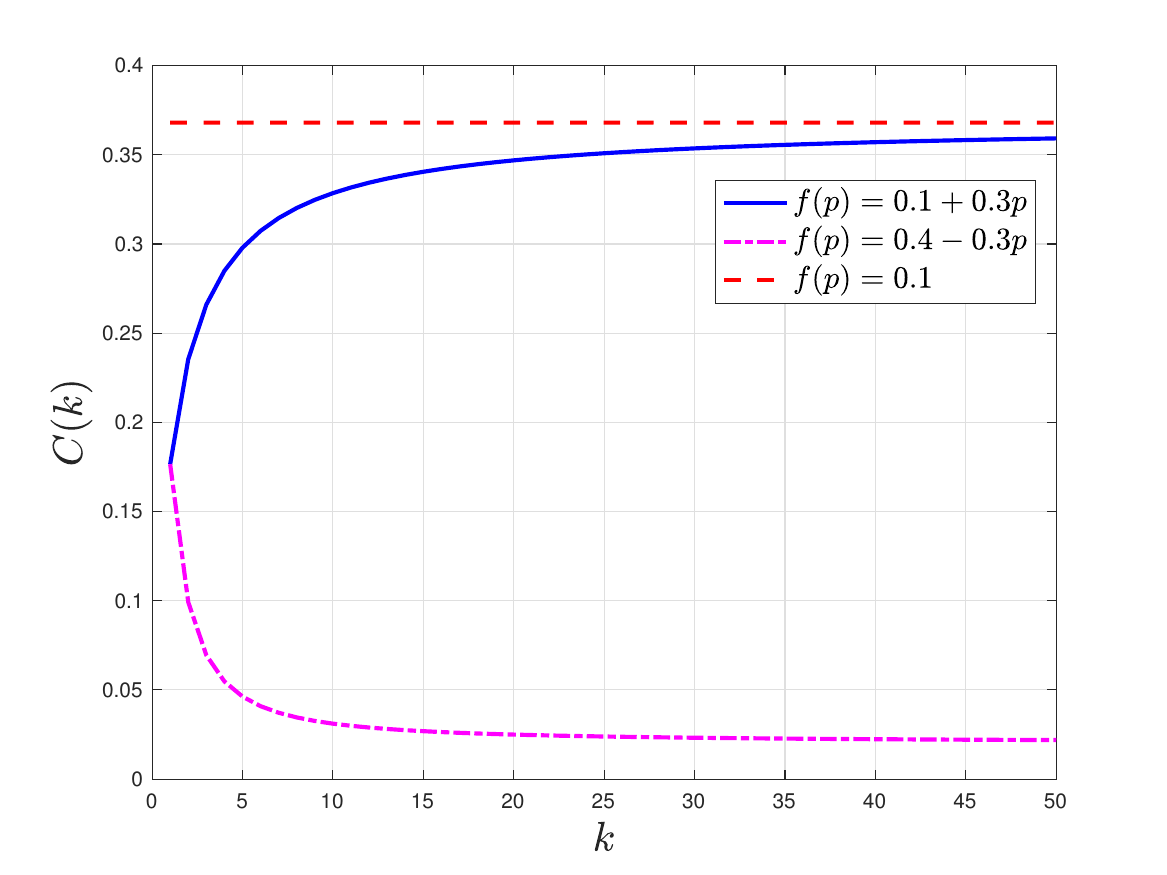}&
\includegraphics[width=.45\columnwidth]{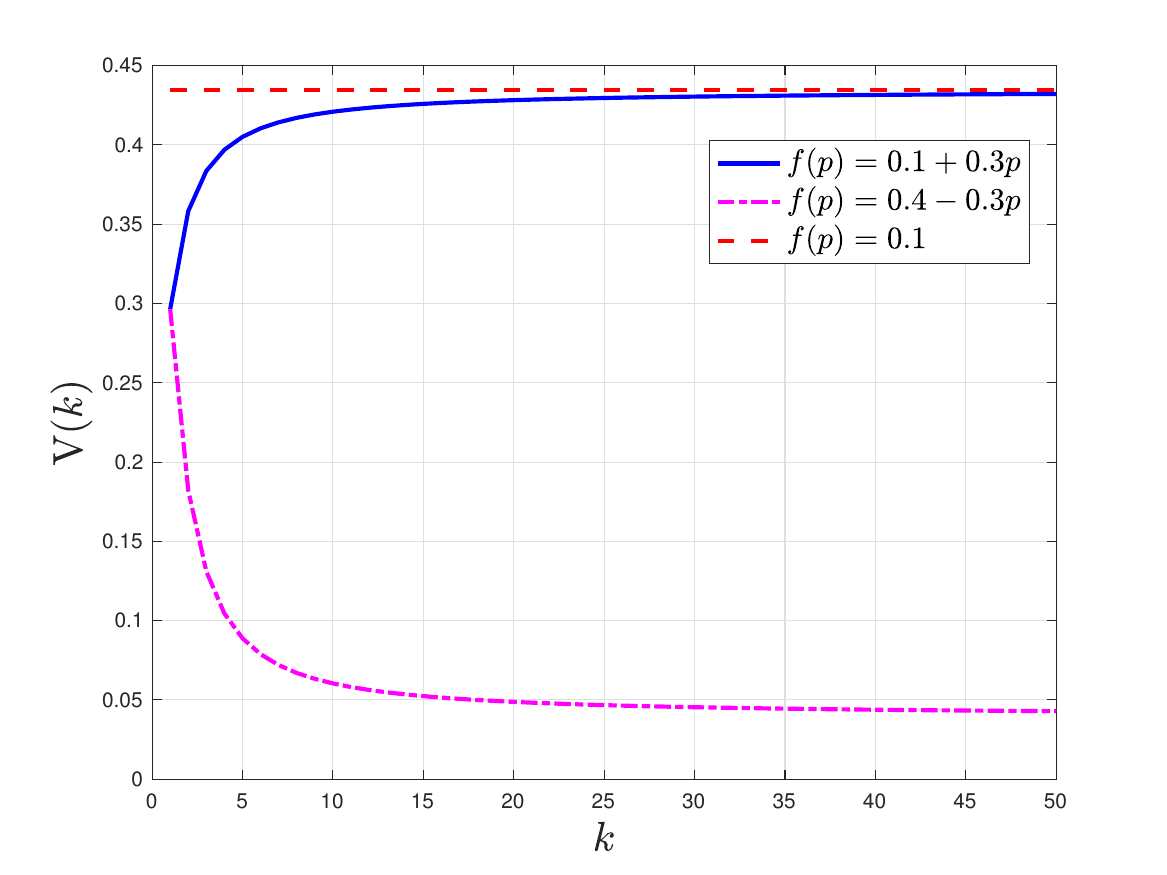}
\end{tabular}
\caption{Plot of $C(k)$ and $\rmV(k)$ for a query-dependent BSC for various values of $k$ and different Lipschitz continuous channel functions $f(\cdot)$.}
\label{capacity_bsc}
\end{figure}

Let $P_{Y|Z}^{f(p)}$ be the query-dependent BSC defined in Def. \ref{def:mdBSC}.  Given any $p\in(0,1)$ and $k\in\bbN$, for any Lipschitz continuous function $f(\cdot)$, let $\beta(p,k):=1-f(p)+(1-p)^k(2f(p)-1)$. In this case, for any $(x_{[k]},y)\in\{0,1\}^{k+1}$, the information density is
\begin{align}
\imath_\emptyset^{f(p),k}(x_{[k]};y)
&=\bbo\{y\neq z\}\log(f(p))+\bbo\{y=z\}\log(1-f(p))-\bbo\{y=1\}\log(\beta(p,k))-\bbo\{y=0\}\log(1-\beta(p,k)),
\end{align}
where $z$ is the binary OR of $x_{[k]}$. The mean and the variance of $\imath_\emptyset^{f(p),k}$ are respectively
\begin{align}
C_\emptyset(p,k)&=h_\rmb(\beta(p,k))-h_\rmb(f(p)),\\
\rmV_\emptyset(p,k)&=\mathrm{Var}[\imath_\emptyset^{f(p),k}(X_{[k]};Y)],
\end{align}
where $h_\rmb(p)=-p\log p-(1-p)\log(1-p)$ is the binary entropy function. The capacity $C(k)$ is then given by
\begin{align}
C(k)=\max_{p\in[0,1]} C_\emptyset(p,k)=\max_{p\in[0,1]} \big(h_\rmb(\beta(p,k))-h_\rmb(f(p))\big).
\end{align}
The variance $\rmV(k,\varepsilon)$ can be calculated using the capacity achieving parameters $p^*$ that achieve $C(k)$.  When $p^*$ is unique, $\rmV(k,\varepsilon)$ is independent of $\varepsilon$ and we write it as $\rmV(k)$ for simplicity. We plot $C(k)$ and $\rmV(k)$ for various  Lipschitz continuous channel functions $f(\cdot)$ in Figure \ref{capacity_bsc}. Note that for a query-independent BSC having $f(p)$ equal to a constant $\alpha$, the capacity $C(k)$ is a constant function of $k$ since $C(k)=\max_{p\in[0,1]}h_\rmb(\beta(p,k))-h_\rmb(\alpha)=h_\rmb(\frac{1}{2})-h_\rmb(\alpha)$ and it is achieved by the $p^*$ such that $\beta(p^*,k)=\frac{1}{2}$, which simplifies to $p^*$ such that $(1-p^*)^k=\frac{1}{2}$.

\subsubsection{Case of a query-Dependent BEC}
\begin{figure}[tb]
\centering
\begin{tabular}{cc}
\includegraphics[width=.45\columnwidth]{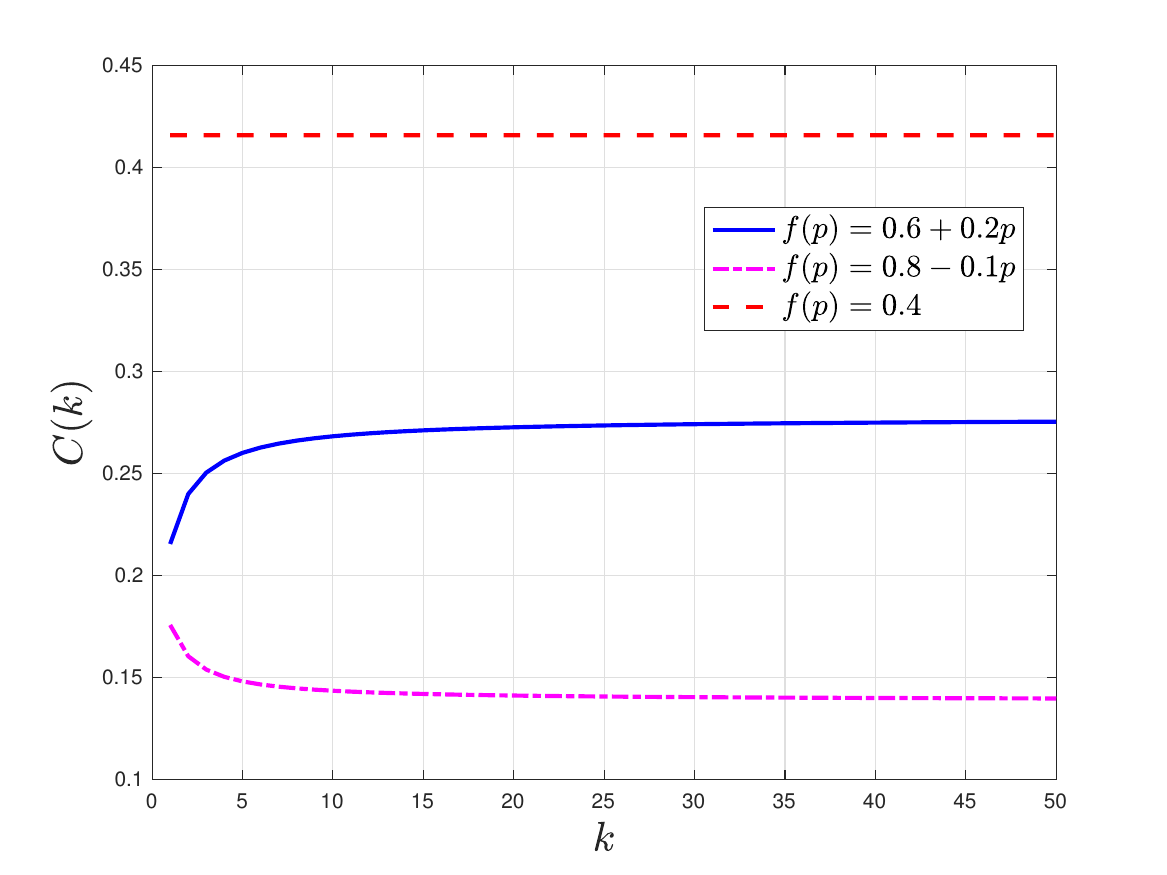}&\includegraphics[width=.45\columnwidth]{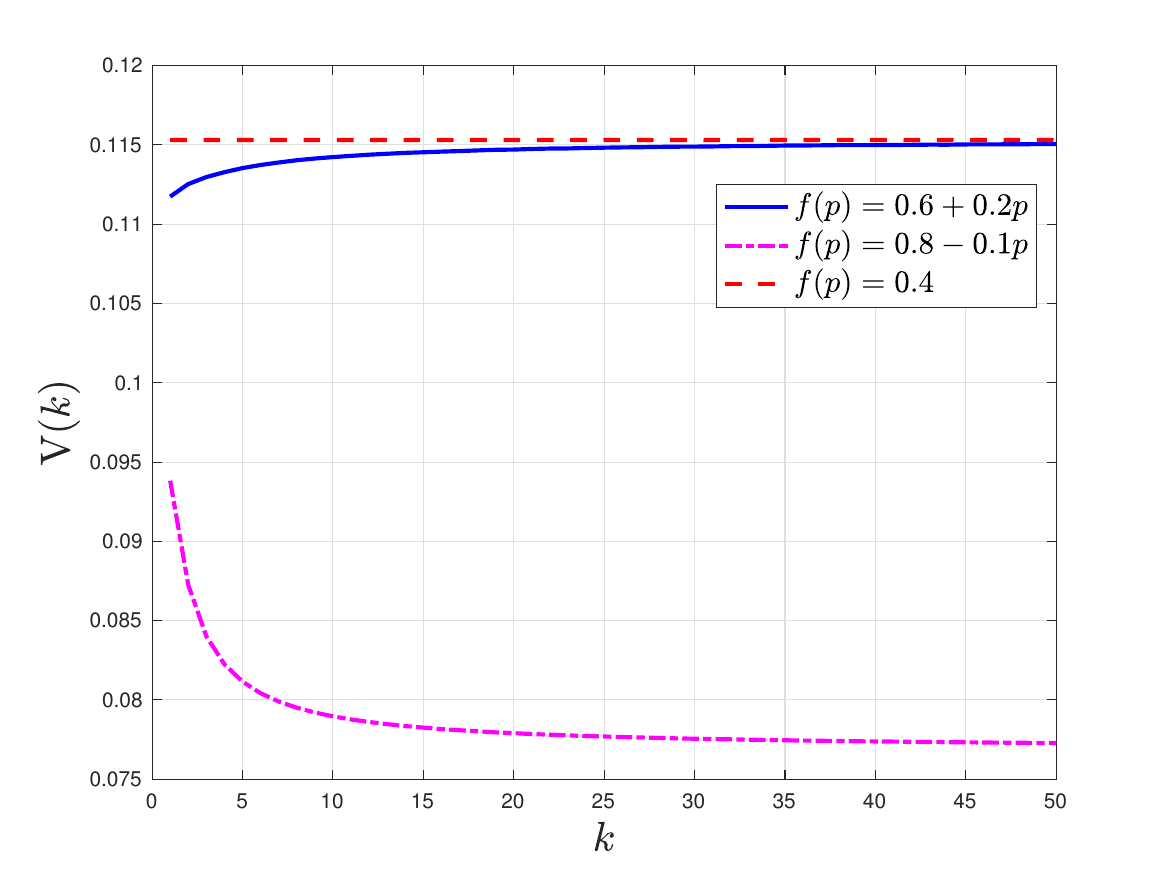}
\end{tabular}
\caption{Plot of $C(k)$ and $\rmV(k)$ for a query-dependent BEC for various values of $k$ and different Lipschitz continuous channel functions $f(\cdot)$.}
\label{capacity_bec}
\end{figure}

We consider the query-dependent BEC $P_{Y|Z}^{f(p)}$ in Def. \ref{def:mdBEC}. In this case, for any $(x_{[k]},y)\in\{0,1\}^k\times\{0,1,\rme\}$, the information density is
\begin{align}
\imath_\emptyset^{f(p),k}(x_{[k]};y)
&=\bbo\{z=0,y=0\}\log\frac{1}{(1-p)^k}+\bbo\{z=1,y=1\}\log\frac{1}{1-(1-p)^k},
\end{align}
where $z$ is the binary OR of $x_{[k]}$. The expectation and the variance of the information density satisfy 
\begin{align}
C_\emptyset(p,k)&=(1-f(p))h_\rmb((1-p)^k),\\
\rmV_\emptyset(p,k)&=\mathrm{Var}[\imath_\emptyset^{f(p),k}(X_{[k]};Y)].
\end{align}
The capacity $C(k)$ is then given by
\begin{align}
C(k)=\max_{p\in[0,1]}(1-f(p))h_\rmb((1-p)^k).
\end{align}
When the capacity achieving parameter $p^*$ is unique, the variance $\rmV(k,\varepsilon)$ is independent of $\varepsilon\in(0,1)$ and we write it as $\rmV(k)$. We plot $C(k)$ and $\rmV(k)$ for various Lipschitz continuous channel functions $f(\cdot)$ in Figure \ref{capacity_bec}. Note, as in the BSC, for a query-independent BEC having $f(p)$ equal to a constant $\alpha$, the capacity $C(k)$ is a constant function of $k$ since $C(k)=(1-\alpha)\max_{p\in[0,1]}h_\rmb((1-p)^k)=(1-\alpha)h_\rmb(\frac{1}{2})$ and it is achieved by the $p^*$ such that $(1-p^*)^k=\frac{1}{2}$. 

The reader will note from Figures \ref{capacity_bsc} and \ref{capacity_bec} that the query-dependent BSC and BEC have the capacities $C(k)$ and the variances $\rmV(k)$ that appear monotonic in the number of targets $k$. This monotonicity mirrors the monotonicity of the Lipschitz continuous channel function $f(p)$. Theoretical characterization of monotonicity does not appear straightforward and is left for future work.

\subsection{Numerical Simulations}
\label{sec:simulation}
\begin{figure}[tb]
\centering
\begin{tabular}{cc}
\includegraphics[width=.45\columnwidth]{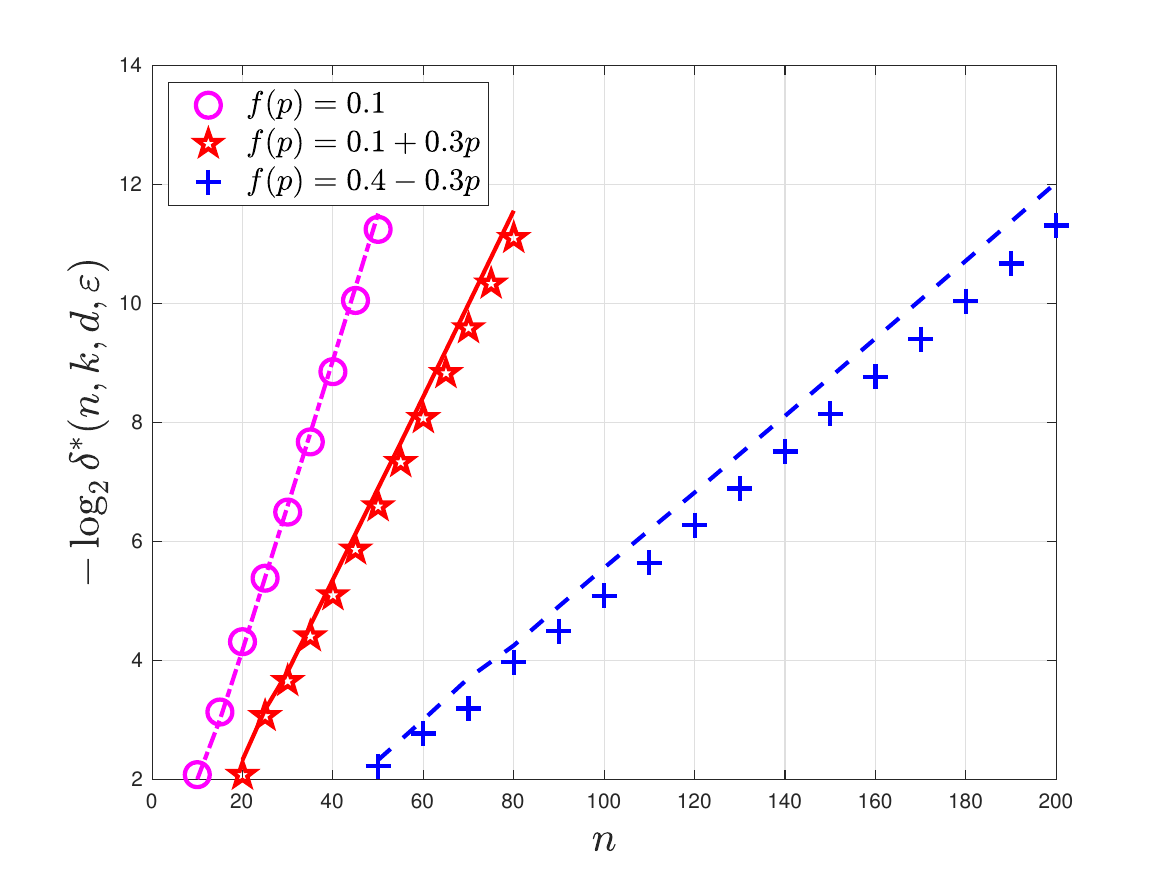}&\includegraphics[width=.45\columnwidth]{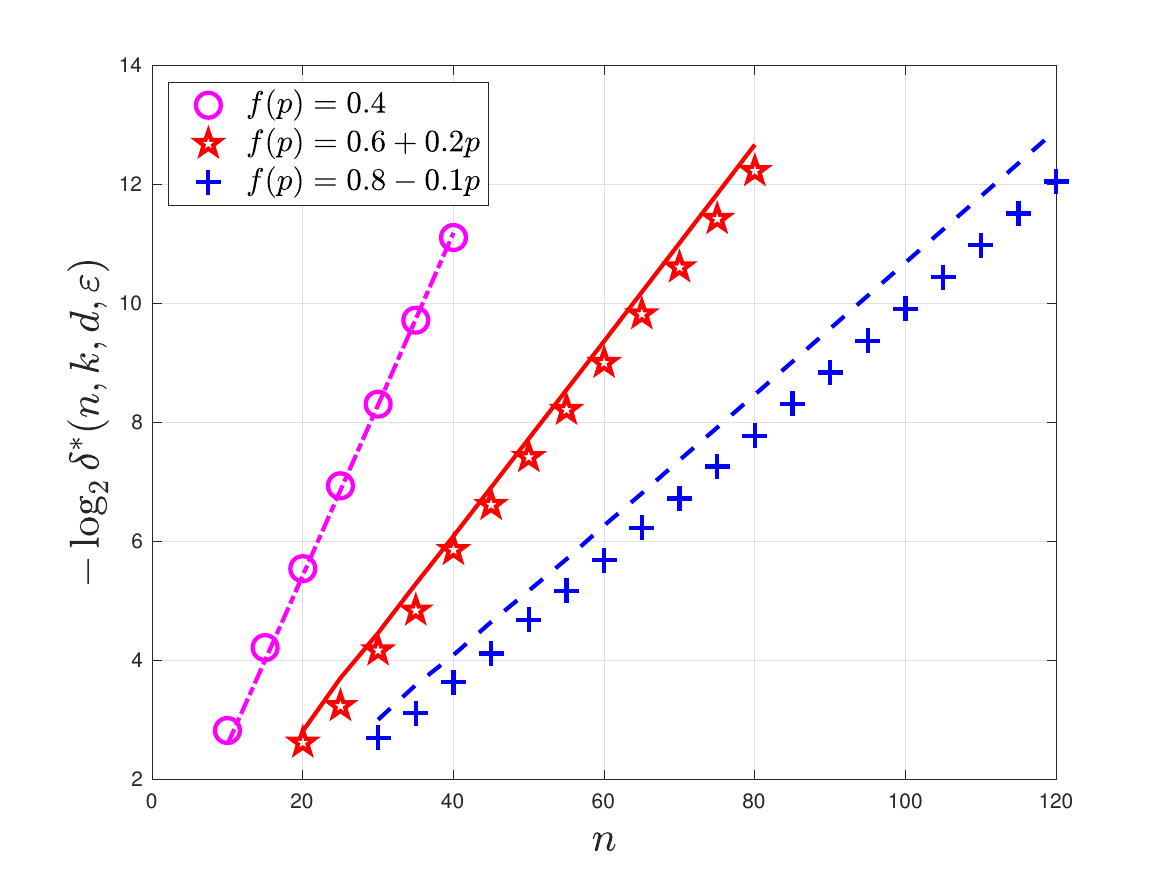}\\
{(a) BSC}&{(b) BEC }
\end{tabular}
\caption{Minimal achievable resolution of the non-adaptive query procedure in Algorithm \ref{procedure:nonadapt:ktaget} for estimating $k=2$ independent one-dimensional target variables $(S_1,S_2)$ in the unit interval with target excess-resolution probability $\varepsilon=0.3$. We consider various Lipschitz continuous functions $f(\cdot)$.
 The solid lines correspond to the second-order asymptotic result in Theorem \ref{result:second:ktarget} and the symbols correspond to the Monte Carlo simulation of Algorithm \ref{procedure:nonadapt:ktaget}.}
\label{sim_non_adap_ktarget}
\end{figure}

To illustrate the tightness of our second-order asymptotic result in Theorem \ref{result:second:ktarget}, we numerically simulate the performance of Algorithm \ref{procedure:nonadapt:ktaget} and compare the simulated result versus the theoretical one, for query-dependent BSC and BEC studied in Section \ref{sec:special}.  We consider the simple case of $k=2$ and $d=1$\footnote{The complexity of Algorithm \ref{procedure:nonadapt:ktaget} is at least $M^d \choose k$, which scales significantly with both $d$ and $k$ for a relatively large $M$. For example, when $M=50$, $d=4$ and $k=2$, we have ${M^d \choose k}=1.9531*10^{13}$, which is well out of the simulation power of a usual computer.}. We assume that the target random variables $(S_1,S_2)$ are independently and uniformly distributed over the set $[0,1]$. In Figure \ref{sim_non_adap_ktarget}, the simulated achievable resolution of our non-adaptive query procedure in Algorithm \ref{procedure:nonadapt:ktaget} is plotted and compared to the second-order asymptotic result in Theorem \ref{result:second:ktarget}.

In our simulation, for each number of queries $n\in\bbN$, the parameter $M$ is chosen such that
\begin{align}
\log M=\frac{nC(2)+\sqrt{nV(2,\varepsilon)}\Phi^{-1}(\varepsilon)-\frac{1}{2}\log n}{2}.
\end{align}
We set $\gamma=\frac{1}{2}\log n$ and choose $p=p^*$, which is a capacity achieving parameter. For each number of queries $n$, the non-adaptive procedure in Algorithm \ref{procedure:nonadapt:ktaget} is run independently $10^4$ times and the achievable resolution is then calculated\footnote{Since we calculate the achievable resolution with respect to a given excess-resolution probability $\varepsilon$, a data point is calculated from $10^3$ runs of the algorithm. We sort the resolutions of all $10^3$ runs and calculate the achievable resolution as the $(1-\varepsilon)\times 10^3$-th smallest resolution. We then repeat the procedure 10 times to obtain the average achievable resolution.}. Figure \ref{sim_non_adap_ktarget} empirically confirms that Theorem \ref{result:second:ktarget} provides a tight approximation to the non-asymptotic performance of Algorithm \ref{procedure:nonadapt:ktaget} under various Lipschitz continuous channel functions $f(\cdot)$.

\section{Proof of the Non-Asymptotic Achievability Bound (Theorem \ref{theorem:fbl:achievability})}
\label{proof:ach:fbl}

\subsection{Preliminaries}
Recall the definitions of the joint distribution $P_{X_{[k]}Y}^{f(p),k}$ in \eqref{def:pjointk}, the information density $\imath_\calJ^{f(p),k}$ in \eqref{def:cdmi} and the (conditional) mutual information $C_\calJ(p,t)$ in \eqref{def:cjpt}. Define the third absolute moment of $\imath_\calJ^{f(p),k}$ as
\begin{align}
T_\calJ(p,t)&:=\mathbb{E}_{P_{X_{[t]}Y}^{f(p),t}}\big[\big|\imath_\calJ^{f(p),t}(X_{[t]};Y)-C_\calJ(p,t)\big|^3\big]\label{def:tjpt}.
\end{align}
Furthermore, given any $(d,M)\in\bbN^2$, define a function $\Gamma:[M]^d\to [M^d]$ such that for any $(i_1,\ldots,i_d)\in[M]^d$,
\begin{align}
\Gamma(i_1,\ldots,i_d)=1+\sum_{j\in[d]}(i_j-1)M^{d-j}\label{def:Gamma}.
\end{align}
Note that the function $\Gamma(\cdot)$ is invertible. We denote $\Gamma^{-1}:[M^d]\to [M]^d$ as the inverse function.

\subsection{Detailed Description of the Non-adaptive Query Procedure}

Consider any integer $M\in\bbN$. Partition the unit cube $[0,1]^d$ into $M^d$ equal-sized disjoint cubes $\calS_1,\ldots,\calS_{M^d}$ and consider any $M^d$ binary vectors $\bx=(x^n(1),\ldots,x^n(M^d))$. For each $l\in[n]$, the $l$-th query is designed as
\begin{align}
\calA_l
&:=\bigcup_{j\in[M^d]:x_l(j)=1}\calS_i\label{def:query:ddim},
\end{align}
For subsequent analysis, given any $s\in[0,1]$, define the following quantization function
\begin{align}
\rmq(s):=\lceil sM\rceil\label{def:qs},
\end{align} 

Suppose we have $k$ targets with location vectors $\bs^k:=(\bs_1,\ldots,\bs_k)$, where for each $i\in[k]$, the $d$-dimensional location vector of the $i$-th target $\bs_i$ is given by $(s_{i,1},\ldots,s_{i,d})$. For each $(i,j)\in[k]\times[d]$, let $w_{i,j}(\bs^k)=\rmq(s_{i,j})$ denote the quantized index of $s_{i,j}$ and let $\bw_i(\bs^k)=(w_{i,1}(\bs^k),\ldots,w_{i,d}(\bs^k))$ denote the quantized indices of $\bs_i$. Recall the definition of $\Gamma(\cdot)$ in \eqref{def:Gamma}. We number\footnote{This can be done as follows. We partition the unit cubes into equal-sized sub-cubes $\{\calB_{(i_1,\ldots,i_d)}\}_{(i_1,\ldots,i_d)\in[M^d]}$. The partition is done by partitioning each dimension into equal-length intervals where the each interval only includes the left end points except the last one that also includes the right end points. We then take the formed sub-cubes as the partition where the index $i_j$ denotes the index of the partition in the $j$-th dimension for each $j\in[d]$. Our proposed partition is formed by choosing $\calS_i=\calB_{\Gamma^{-1}(i)}$ for each $i=[M^d]$.} the sub-cubes so that for each $i\in[k]$, the $i$-th target with location vector $\bs_i$ lies in the sub-cube with index $\calS_{\Gamma(\bw_i(\bs^k))}$.

It is possible that two targets are quantized into the same sub-cube, i.e., there exists a pair of indices $(i,j)\in[k]^2$ such that $i\neq j$ and $\bw_i=\bw_j$. If so, the detected number of targets would be smaller than $k$. To account for these cases, we define the set of unique quantized indices as
\begin{align}
\calW_\rmp(\bs^k):=\{w\in[M^d]:~\exists~i\in[k],~\Gamma(\bw_i(\bs^k))=w\}\label{def:wp},
\end{align}
and define the number of distinct quantized targets as
\begin{align}
k_\rmp(\bs^k):=|\calW_\rmp(\bs^k)|\label{def:kp}.
\end{align}
Furthermore, let $w^{\uparrow}_\rmp(\bs^k)$ be the vector that orders elements in $\calW_\rmp(\bs^k)$ increasingly and let $w^{\uparrow}_\rmp(\bs^k,i)$ be the $i$-th element. It follows that $w^{\uparrow}_\rmp(\bs^k)\in\calL(k,M^d)$ (cf. \eqref{def:calL}). For each $l\in[n]$, the noiseless answer to the query $\calA_l$ (cf. \eqref{def:query:ddim}) is
\begin{align}
z_l
:&=\bbo\{\exists~i\in[k]:~\bs_i\in\calA_l\}\\
&=1\bigg\{\exists~i\in[k]:~\bs_i\in\bigcup_{j\in[M^d]:x_l(j)=1}\calS_i\bigg\}\\
&=\bbo\{\exists~i\in[k_\rmp(\bs^k)]:~x_l(w^{\uparrow}_\rmp(\bs^k,i))=1\}.
\end{align}
The noisy response $Y_l$ is obtained by passing the noiseless response $z_l$ through the query-dependent channel $P_{Y|Z}^{\calA_l}$. 

Recall the definition of $\calD^{n,t}(\gamma)$ in \eqref{def:calD} for each $t\in[k]$. Given noisy responses $Y^n=(Y_1,\ldots,Y_n)$, if there exists a tuple $(i_1,\ldots,i_k)\in\calL(k,M^d)$ such that $(x^n(i_1),\ldots,x^n(i_k),Y^n)\in\calD^{n,k}(\gamma)$, the decoder $g$ first produces estimates of quantized indices $(\hat{\bw}_1,\ldots,\hat{\bw}_k)$ where for each $i\in[k]$, $\hat{\bw}_i=(\hatw_{i,1},\ldots,\hatw_{i,d})=\Gamma^{-1}(i_k)$. Subsequently, the decoder $g$ outputs the set of estimated location vectors $\calS_m$ as the centers of sub-cubes $\calS_{i_1},\ldots,\calS_{i_k}$, i.e., $\calS_m=\{\hat{\bs}_1,\ldots,\hat{\bs}_k\}$ with 
\begin{align}
\hat{\bs}_{i,j}=\frac{2\hatw_{i,j}-1}{2M},
\end{align}
where $\hat{\bs}_{i,j}$ is the $j$-th element of the $d$-dimensional vector $\hat{\bs}_i$ for each $(i,j)\in[k]\times[d]$. If no such tuple exists, then the decoder $g$ sets $t=k-1$ and uses $t$ in the role of $k$ to continue the previous steps until $t=0$, as described in Algorithm \ref{procedure:nonadapt:ktaget}. Note that if no such tuple exists for all $t\in[k]$, the reproduced set $\calS_m$ is empty and an error is declared.

\subsection{Analysis of Excess-Resolution Events}
\label{sec:events}
Given the above non-adaptive query procedure (see also Algorithm \ref{procedure:nonadapt:ktaget}) using binary query vectors $\bx=(x^n(1),\ldots,x^n(M^d))$, an excess-resolution event for target location vectors $\bs^k=(\bs_1,\ldots,\bs_k)$ occurs if at least one of the following three events occurs:
\begin{itemize}
\item $\calE_1(\bs^k,\bx,Y^n):$\\
$(x^n(w^\uparrow_\rmp(\bs^k,1)),\!\ldots\!,x^n(w^\uparrow_\rmp(\bs^k,k_\rmp(\bs^k)),Y^n)\!\notin\calD^{n,k_\rmp(\bs^k)}(\gamma)$;
\item $\calE_2(\bs^k,\bx,Y^n)$: there exists a tuple $(i_1,\ldots,i_{k_\rmp(\bs^k)})\in\calL(k_\rmp(\bs^k),M^d)$ such that $(x^n(i_1),\ldots,x^n(i_{k_\rmp(\bs^k)}),Y^n)\in\calD^{n,k_\rmp(\bs^k)}(\gamma)$ and $(i_1,\ldots,i_{k_\rmp(\bs^k)})\neq w^\uparrow_\rmp(\bs^k)$;
\item $\calE_3(\bs^k,\bx,Y^n):$ for some $t\in[k_\rmp(\bs^k)+1:k]$, there exists a tuple $(i_1,\ldots,i_t)\in\calL(t,M^d)$ such that $(x^n(i_1),\ldots,x^n(i_t),Y^n)\in\calD^{n,t}(\gamma)$.
\end{itemize}
Note that $\calE_1$ denotes the event where the true quantized index vector fails the information density threshold decoding rules. The error event $\calE_2$ denotes the event where a quantized index vector different from the true quantized index vector, but with the same length, satisfies the threshold rules. The event $\calE_3$ denotes the event where a quantized index vector, having a length different from the true quantized index vector, satisfies the threshold rules. Similar information spectrum definitions were used in the finite blocklength analysis of MAC~\cite{molavianjazi2015second} and RAC~\cite{yavas2018random}.

Thus, using the non-adaptive query procedure in Algorithm \ref{procedure:nonadapt:ktaget}, given any target location vectors $\bs^k$, any binary query vectors $\bx$ and any $M\in\bbN$, the excess-resolution probability with respect to the resolution level $\frac{1}{M}$ satisfies
\begin{align}
\rmP_\rme(\bs^k,\bx)
&:=\max\bigg\{\Pr\Big\{\exists~i\in[k]:~\min_{\hat{\bs}\in\calS_m}|\hat{\bs}-\bs_i|_{\infty}>\frac{1}{M}\Big\},\Pr\Big\{\exists~\hat{\bs}\in\calS_m:~\min_{i\in[k]}\|\hat{\bs}-\bs_i\|_{\infty}>\frac{1}{M}\Big\}\bigg\}\\
&\leq \Pr\Big\{\bigcup_{i\in[3]}\calE_i(\bs^k,\bx,Y^n)\Big\}\\
&\leq \sum_{i\in[3]}\Pr\{\calE_i(\bs^k,\bx,Y^n)\}\label{upp:psx},
\end{align}
where the probability terms are calculated with respect to the conditional distribution of the noisy responses $Y^n$ given $\bx$ and $\bs^k$, which is induced by the query-dependent channel.

\subsection{Employ Random Query Vectors and Change of Measure}
For subsequent analysis, we will employ a random coding argument where we consider random binary query vectors $\bX:=(X^n(1),\ldots,X^n(M^d))$. For each $i\in[M^d]$, the length-$n$ vector $X^n(i)$ is generated i.i.d. from $P_X$, a Bernoulli distribution with parameter $p\in(0,1)$. Given any $\bs^k$, the joint distribution of $(\bX,Y^n)$ under queries $\calA^n=(\calA_1,\ldots,\calA_n)$ is
\begin{align}
P_{\bX Y^n}^{\calA^n}(\bx,y^n)
&=\Big(\prod_{j\in[M^d]}P_X^n(x^n(j))\Big)\times\Big(\prod_{l\in[n]}P_{Y|Z}^{\calA_l}(y_l|z_l)\Big)\label{truechannel},
\end{align}
where we use $z_l$ to denote $\bbo\{\exists~i\in[k_\rmp(s^k)]:~x_l(w_\rmp^\uparrow(\bs^k,i))=1\}$.

For any $\eta\in\bbR_+$, define the following typical set of query vectors:
\begin{align}
\calT^n(M,d,p,\eta)
:=\Big\{\bx=(x^n(1),\ldots,x^n(M^d))\in([0,1]^n)^{M^d}:~\max_{t\in[n]}|q_{t,d}^{M,n}(\bx)-p|\leq \eta\Big\},
\end{align}
where 
\begin{align}
q_{t,d}^M(\bx)&:=\frac{1}{M^d}\sum_{j\in[M^d]}x_t(j).
\end{align}
Similarly to the proof of \cite[Theorem \ref{theorem:fbl:achievability}]{zhou2019twentyq}, we have that for any $\eta>0$,
\begin{align}
\mathbb{E}[\rmP_\rme(\bs^k,\bX)]
&\leq \mathbb{E}[\rmP_\rme(\bs^k,\bX)\bbo\{\bX\in\calT^n(M,d,p,\eta)\}]+\Pr\{\bX\notin\calT^n(M,d,p,\eta)\}\label{ach:move2pre}\\
&\leq \mathbb{E}[\rmP_\rme(\bs^k,\bX)\bbo\{\bX\in\calT^n(M,d,p,\eta)\}]+4n\exp(-2M^d\eta^2)\label{ach:move2},
\end{align}
where \eqref{ach:move2} follows from \cite[Lemma 22]{tan2014state}, which upper bounds the second term in \eqref{ach:move2pre} using the union bound and Hoeffding‘s inequality. 

In the rest of this section, we upper bound the first term in \eqref{ach:move2}. We will need the following query-independent distribution $P_{\bX Y^n}^{f(p)}$ given any $\bs^k$:
\begin{align}
P_{\bX Y^n}^{f(p)}(\bx,y^n)
&=\Big(\prod_{j\in[M^d]}P_X^n(x^n(j))\Big)\times\Big(\prod_{l\in[n]}P_{Y|Z}^{f(p)}(y_l|z_l)\Big)\label{def:pmi},
\end{align}
where $z_l$ is the same as in \eqref{truechannel}. Using the assumption on the query-dependent channel in \eqref{assump:continuouschannel}, for any $\bx\in\calT^n(M,d,p,\eta)$,
\begin{align}
\log \frac{P_{\bX Y^n}^{\calA^n}(\bx,y^n)}{P_{\bX Y^n}^{f(p)}(\bx,y^n)}
&=\sum_{l\in[n]}\log\frac{P_{Y|Z}^{f(q_{t,d}^M(\bx))}(y_l|z_l)}{P_{Y|Z}^{f(p)}(y_l|z_l)}\leq n\mu\eta c(f(p))\label{cofmeasure},
\end{align}
where \eqref{cofmeasure} follows since $\bx\in\calT^n(M,d,p,\eta)$ implies that $|q_{t,d}^{M,n}(\bx)-p|\leq \eta$ and the Lipschitz continuous property of the function $f$ ensures that $|f(q_1)-f(q_2)|\leq \mu |q_1-q_2|$ for any $(q_1,q_2)\in[0,1]^2$.

Combining \eqref{upp:psx} and \eqref{cofmeasure}, we have
\begin{align}
\nn&\mathbb{E}[\rmP_\rme(\bs^k,\bX)\bbo\{\bX\in\calT^n(M,d,p,\eta)\}]\\*
&\leq \Pr\{\cup_{i\in[3]}\calE_i(\bs^k,\bX,Y^n)\bbo\{\bX\in\calT^n(M,d,p,\eta)\}\}\\
&\leq \exp(n\mu\eta c(f(p)))\Pr_{P_{\bX Y^n}^{f(p)}}\{\cup_{i\in[3]}\calE_i(\bs^k,\bX,Y^n)\}\\
&\leq \exp(n\mu\eta c(f(p)))\bigg(\Pr_{P_{\bX Y^n}^{f(p)}}\{\calE_1(\bs^k,\bX,Y^n)\}+\Pr_{P_{\bX Y^n}^{f(p)}}\{\calE_2(\bs^k,\bX,Y^n)\}+\Pr_{P_{\bX Y^n}^{f(p)}}\{\calE_3(\bs^k,\bX,Y^n)\}\bigg)\label{ach:step2}.
\end{align}

\subsection{Final Steps}
\label{sec:derivebound}
From the definition of excess-resolution events in Section \ref{sec:events}, the first term in the parenthesis of \eqref{ach:step2} is
\begin{align}
\Pr_{P_{\bX Y^n}^{f(p)}}\{\calE_1(\bs^k,\bX,Y^n)\}
=\Pr_{P_{\bX Y^n}^{f(p)}}\Big\{(X^n(w^\uparrow_\rmp(\bs^k,1)),\ldots,X^n(w^\uparrow_\rmp(\bs^k,k_\rmp(\bs^k)),Y^n)\notin\calD^{n,k_\rmp(\bs^k)}(\gamma)\Big\}.
\end{align}
We then bound the remaining probability terms inside the parenthesis of \eqref{ach:step2}.

For subsequent analysis, define the following set for each $j\in[k_\rmp(\bs^k)]$:
\begin{align}
\calI_j(\bs^k)
&:=\Big\{(i_1,\ldots,i_{k_\rmp(\bs^k)})\in\calL(k_\rmp(\bs^k),M^d):(i_1,\ldots,i_{k_\rmp(\bs^k)})\neq w^{\uparrow}_\rmp(\bs^k)\mathrm{~and~}\big|\{l\in[k_\rmp(\bs^k)]:~i_l\neq w^{\uparrow}_\rmp(\bs^k,l)\}\big|=j\Big\}\label{def:calI_j}.
\end{align}
Note that $\calI_j(\bs^k)$ denotes the set of length-$k_\rmp(\bs^k)$ vectors that are different from the quantized indices $w^{\uparrow}_\rmp(\bs^k)$ by exactly $j$ elements. For any $\bi:=(i_1,\ldots,i_{k_\rmp(\bs^k)})\in\calI_j(\bs^k)$, we use $X^n_{\bi}$ to denote the random binary vector $(X^n(i_1),\ldots,X^n(i_{k_\rmp(\bs^k)}))$. The joint distribution of $(X^n(\bi_{\bs^k}),Y^n)$ are induced by the joint distribution $P_{\bX Y^n}^{f(p)}$ in \eqref{def:pmi}.  Furthermore, we define a set
\begin{align}
\calJ(\bi,\bs^k):=\{t\in[k_\rmp(s^k)]:i_t\in\calW_\rmp(\bs^k)\}.
\end{align}
Then we use $X^n_{\calJ(\bi,\bs^k)}$ to denote the collection of $X^n(j)$ with $j\in\calJ(\bi,\bs^k)$ and we use $x_{\calJ(\bi,\bs^k)}^n$, $X_{\calJ(\bi,\bs^k)}$ and $x_{\calJ(\bi,\bs^k)}$  similarly. Note that under the joint distribution $P_{\bX Y^n}^{f(p)}$ in \eqref{def:pmi}, the noisy responses $Y^n$ depend on $X^n_{\bi}$ only through $X^n_{\calJ(\bi,\bs^k)}$, and we use $P_{Y^n|X^n_{\calJ(\bi,\bs^k)}}^{f(p)}$ to denote the induced conditional distribution, i.e.,
\begin{align}
P_{Y^n|X^n_{\bi}}(y^n|x^n(\bi))
&=P_{Y^n|X^n_{\calJ(\bi,\bs^k)}}^{f(p)}(y^n|x^n_{\calJ(\bi,\bs^k)})=\prod_{l\in[n]}P_{Y|X_{\calJ(\bi,\bs^k)}}^{f(p),k_\rmp(s^k)}(y_l|x_{l,\calJ(\bi,\bs^k)}),\label{useprevious}
\end{align}
where \eqref{useprevious} follows from the definition of the joint distribution $P_{\bX Y^n}^{f(p)}$ in \eqref{def:pmi} and $P_{Y|X_\calJ}^{f(p),k}$ is the induced marginal distribution of the joint distribution $P_{X_{[k]}Y}^{f(p),k}$ in \eqref{def:pjointk} for any $\calJ\subseteq[k]$.

Using the information spectrum method introduced in \cite{han2006information} (see also \cite[Lemma 7.10.1]{han2003information}), we upper bound $\Pr\{\calE_2(\bs^k,\bX,Y^n)\}$ as follows:
\begin{align}
\Pr_{P_{\bX Y^n}^{f(p)}}\{\calE_2(\bs^k,\bX,Y^n)\}
&=\!\!\!\!\!\!\sum_{j\in[k_\rmp(\bs^k)]}\Pr_{P_{\bX Y^n}^{f(p)}}\big\{\exists~\bi\in\calI_j(\bs^k):~(X^n(\bi),Y^n)\in\calD^{n,k_\rmp(\bs^k)}(\gamma)\big\}\label{usecalI_j}\\
&\leq\sum_{j\in[k_\rmp(\bs^k)]}\sum_{\bi\in\calI_j(\bs^k)}\Pr_{P_{\bX Y^n}^{f(p)}}\!\!\big\{(X^n(\bi),Y^n\!)\!\in\!\calD^{n,k_\rmp(\bs^k)}(\gamma)\!\big\}\\
&=\!\!\!\!\!\!\sum_{j\in[k_\rmp(\bs^k)]}\sum_{\bi\in\calI_j(\bs^k)}\sum_{\substack{(\bx,y^n):\\(x^n_{\calJ(\bi,\bs^k)},y^n)\in\calD^{n,k_\rmp(\bs^k)}(\gamma)}}\!\!\!\!\!\!P_{\bX Y^n}^{f(p)}(\bx,y^n)\\
&=\sum_{j\in[k_\rmp(\bs^k)]}\sum_{\bi\in\calI_j(\bs^k)}
\sum_{\substack{(x^n(\bi),y^n)\\\in\calD^{n,k_\rmp(\bs^k)}(\gamma)}}\!\!\!
\Big(\prod_{t\in[k_\rmp(\bs^k)]}P_X^n(x^n(i_t))\Big)P_{Y^n|X^n_{\calJ(\bi,\bs^k)}}^{f(p)}(y^n|x^n_{\calJ(\bi,\bs^k)})\\
&\leq\sum_{j\in[k_\rmp(\bs^k)]}\sum_{\bi\in\calI_j(\bs^k)}
\sum_{\substack{(x^n(\bi),y^n)\\\in\calD^{n,k_\rmp(\bs^k)}(\gamma)}}
\Big(\prod_{t\in[k_\rmp(\bs^k)]}P_X^n(x^n(i_t))\Big)P_{Y^n|X^n_{\bi}}^{f(p)}(y^n|x^n_{\bi})\frac{\exp(-\gamma)}{M^{dj}}\label{usedefDj}\\
&=\sum_{j\in[k_\rmp(\bs^k)]}\sum_{(i_1,\ldots,i_{k_\rmp(\bs^k)})\in\calI_j(\bs^k)}\frac{\exp(-\gamma)}{M^{dj}}\\
&\leq \sum_{j\in[k_\rmp(\bs^k)]}{k_\rmp(\bs^k) \choose j}\exp(-\gamma)\label{finalhaha}\\
&\leq 2^{k_\rmp(\bs^k)}\exp(-\gamma)\\
&\leq 2^k\exp(-\gamma),\label{finale2}
\end{align}
where \eqref{usecalI_j} follows since from the definition of $\calI_j(\bs^k)$ in \eqref{def:calI_j} implies that $\cup_{j\in[k_p(\bs^k)]}\calI_j(\bs^k)=\calL(k_p(\bs^k),M^d)\setminus\{w^{\uparrow}_\rmp(\bs^k)\}$,  \eqref{usedefDj} follows from the definition of $\calD^{n,t}(\cdot)$ in \eqref{def:calD}, 
\eqref{finalhaha} follows since the size of $\calI_j(\bs^k)$ is no greater than ${k_\rmp(\bs^k) \choose j}M^{dj}$ and \eqref{finale2} follows since $k_\rmp(\bs^k)\leq k$.

Similarly to steps leading to \eqref{finale2}, we have
\begin{align}
\Pr_{P_{\bX Y^n}^{f(p)}}\{\calE_3(\bs^k,\bX,Y^n)\}
&\leq \sum_{t\in[k_\rmp(\bs^k)+1:k]}2^t\exp(-\gamma)\\
&\leq 2^k\sum_{t\in[k_\rmp(\bs^k)+1:k]}\exp(-\gamma)\\
&=2^k\big(k-k_\rmp(\bs^k)\big)\exp(-\gamma)\\
&\leq k2^k\exp(-\gamma)\label{finale3}.
\end{align}

Combining \eqref{ach:move2}, \eqref{ach:step2}, \eqref{finale2} and \eqref{finale3}, we conclude that for any target location vector $\bs^k=(\bs_1,\ldots,\bs_k)$ and any parameter $M\in\bbN$, the excess-resolution probability of the non-adaptive query procedure in Algorithm \ref{procedure:nonadapt:ktaget} satisfies
\begin{align}
\mathbb{E}[\rmP_\rme(\bs^k,\bX)]
&\leq 4n\exp(-2M^d\eta^2)+\exp(n\mu\eta c(f(p)))\bigg((k+1)2^k\exp(-\gamma)+\Pr_{P_{\bX Y^n}^{f(p)}}\{\calE_1(\bs^k,\bX,Y^n)\}\bigg)\label{ktarget:step1}.
\end{align}
Our subsequent analysis focuses on the probability term in \eqref{ktarget:step1}. Consider any $\bs^k$ such that $k_\rmp(\bs^k)=t$ for some $t\in[k]$ and let $w^{\uparrow}_\rmp(\bs^k)=[t]$. Recall the definitions of the (conditional) mutual information density in \eqref{def:cdmi}
and its statistics in \eqref{def:cjpt} to \eqref{def:tjpt}. It follows that
\begin{align}
\Pr_{P_{\bX Y^n}^{f(p)}}\{\calE_1(\bs^k,\bX,Y^n)\}
&=\Pr_{P_{\bX Y^n}^{f(p)}}\{(X^n_{[t]},Y^n)\notin\calD^{n,t}(\gamma)\}\\
&=\Pr_{P_{\bX Y^n}^{f(p)}}\Bigg\{(X^n_{[t]},Y^n)\notin\bigcap_{\calJ\subset[t]}\calD_\calJ^{n,t}(\gamma)\Big\}\\
&\leq \max_{t\in[k]}\Pr_{P_{\bX Y^n}^{f(p)}}\Big\{(X^n_{[t]},Y^n)\notin\bigcap_{\calJ\subset[t]}\calD^{n,t}_{\calJ}(\gamma)\Big\}\label{upplastterm}.
\end{align}

Combining \eqref{ktarget:step1} and \eqref{upplastterm} leads to 
\begin{align}
\mathbb{E}[\rmP_\rme(\bS^k,\bX)]
\leq 4n\exp(-2M^d\eta^2)+\exp(n\mu\eta c(f(p)))\bigg((k+1)2^k\exp(-\gamma)+\max_{t\in[k]}\Pr_{(P_{X_{[t]} Y}^{f(p),t})^n}\Big\{(X^n_{[t]},Y^n)\notin\bigcap_{\calJ\subset[t]}\calD_{\calJ}^{n,t}(\gamma)\Big\}\bigg)\label{ktarget:step2},
\end{align}
where \eqref{ktarget:step2} follows from the definition of the joint distributions of $P_{X_{[k]}Y}^{f(p),k}$ in \eqref{def:pjointk} and $P_{\bX Y^n}^{f(p)}$ in \eqref{def:pmi}.

Finally, the existence of deterministic binary vectors $\bx=(x^n(1),\ldots,x^n(M^d))$ with desired performance is guaranteed by the simple fact that $\mathbb{E}[X]\leq a$ implies that there exists $x\leq a$ for any random variable $X$ and real number $a$.

\section{Proof of the Non-Asymptotic Converse Bound (Theorem \ref{theorem:fbl:converse})}
\label{proof:fbl:con}

\subsection{Preliminaries}
\label{sec:preconverse}
For smooth presentation of our proof, we give necessary definitions in this subsection. For any $\beta\leq\frac{\varepsilon}{2}\leq 0.5$, let $\tilM:=\lfloor \frac{\beta}{\delta}\rfloor$. Similar to \eqref{def:qs}, define the quantization function $\rmq_\beta(\cdot)$:
\begin{align}
\rmq_\beta(s)=\lceil s\tilM\rceil .
\end{align}
Given any $\bs^k$, for each $(i,j)\in[k]\times[d]$, let $w^\beta_{i,j}(\bs^k):=\rmq_\beta(s_{i,j})$ and let $\bw_i^\beta(\bs^k)$ denote $(w_{i,1}^\beta(\bs^k),\ldots,w_{i,d}^\beta(\bs^k))$. For a random vector $\bS^k$, we use $W_{i,j}^\beta(\bS^k)$ and $\bW_i^\beta(\bS^k)$ respectively. Given any $m\in[k]$ and any set $\calS_m\subseteq([0,1]^d)^k$, for each $\hat{\bs}\in\calS_m$, we use $\hat{\bw}^\beta(\hat{\bs})$ to denote the quantized vector $(\rmq_\beta(\hats_1),\ldots,\rmq_\beta(\hats_d))$. Analogous to \eqref{def:wp}, define the following sets
\begin{align}
\calW_\beta(\bs^k)
&:=\{\Gamma(\bw_1^\beta(\bs^k)),\ldots,\Gamma(\bw_d^\beta(\bs^k))\},\label{def:calWsk}\\
\hat{\calW}_\beta(\calS_m)
&:=\big\{\Gamma(\hat{\bw}^\beta(\hat{\bs})):~\hat{\bs}\in\calS_m\big\}\label{def:hatcalW}.
\end{align}
where $\Gamma(\cdot)$ was defined in \eqref{def:Gamma}.

For each $\bs^k\in([0,1]^d)^n$, let $k_\rmp^\beta(\bs^k):=|\calW_\beta(\bS^k)|$ denote the number of present targets after quantization. Note that $k_\rmp^\beta(\bs^k)$ can be smaller than $k$ if two or more targets are quantized into the same region, i.e., there exists $(i,j)\in[k]^2$ such that $i\neq j$ and $\bw_i^\beta(\bs^k)=\bw_j^\beta(\bs^k)$. Furthermore, we use $\bw_\beta^{\uparrow}(\bs^k)$ to denote the vector that orders elements in $\calW_\beta(\bS^k)$ (cf. \eqref{def:calWsk}) increasingly and we use $\bw_\beta^{\uparrow}(\bs^k,i)$ to denote the $i$-th element of $\bw_\beta^{\uparrow}(\bs^k)$ for each $i\in[k_\rmp^\beta(\bs^k)]$. For a random vector $\bS^k$, we use $\bW_\beta^{\uparrow}(\bS^k)$ and $\bW_\beta^{\uparrow}(\bS^k,i)$ respectively. For a reproduced target location vector $\hat{\bs}\in\calS_m$, we use ${\hat{\bw}}_\beta^{\uparrow}(\calS_m)$ and ${\hat{\bw}}_\beta^{\uparrow}(\calS_m,i)$ similarly. Analogous to the definition of $\Gamma(\cdot)$ in \eqref{def:Gamma}, we define two functions $\Pi(\cdot)$ such that
\begin{align}
\Pi(\bw_\beta^{\uparrow}(\bs^k))&:=1+\sum_{j\in[k_\rmp^\beta(\bS^k)]}(\bw_\beta^{\uparrow}(\bs^k,j)-1)(M^d)^{k_\rmp^\beta(\bS^k)-j},\\
\Pi({\hat{\bw}}_\beta^{\uparrow}(\calS_m))&=1+\sum_{j\in[m]}(\bW_\beta^{\uparrow}(\bS^k,j)-1)(M^d)^{m-j}.
\end{align}

\subsection{Lower Bound the Excess-Resolution Probability}
Consider any $(n,k,d,\delta,\varepsilon)$-non adaptive query procedure with queries $\calA^n\subseteq([0,1]^d)^n$ and a decoder $g:\calY^n\to\calS_m\subseteq([0,1]^d)^k$. It follows that
\begin{align}
\varepsilon
&\geq \rmP_\rme(n,k,d,\delta)\\
&=\sup_{f_{\bS}\in\calF([0,1]^d)}\max\Big\{\Pr\big\{\exists~i\in[k]:~\min_{\hat{\bs}\in\calS_m}|\hat{\bs}-\bS_i|_{\infty}>\delta\big\},\Pr\big\{\exists~\hat{\bs}\in\calS_m:~\min_{i\in[k]}|\hat{\bs}-\bS_i|_{\infty}>\delta\big\}\Big\}\label{ktarget:step00},
\end{align}
where for each $(i,j)\in[k]\times[d]$, $S_{i,j}$ is the $j$-th element of the $i$-th target location vector $\bS_i$ and $\hats_j$ is the $j$-th element of a vector $\hat{\bs}$ from $\calS_m$, the set of estimated location vectors.

In this subsection, we relate the first probability term in \eqref{ktarget:step00} with the probability of the event that the quantized reproduced set $\hat{\calW}_\beta(\calS_m)$ has the same size as the set of quantized target locations $\calW_\beta(\bS^k)$ but with different elements. Specifically, for any pdf $f_{\bS}$, we have
\begin{align}
\nn&\Pr\{\hat{\calW}_\beta(\calS_m)\neq \calW_\beta(\bS^k)\mathrm{~and~}|\hat{\calW}_\beta(\calS_m)|=|\calW_\beta(\bS^k)|\}\label{errorstart}\\*
\nn&=\Pr\Big\{\hat{\calW}_\beta(\calS_m)\neq \calW_\beta(\bS^k),~|\hat{\calW}_\beta(\calS_m)|=|\calW_\beta(\bS^k)|~\mathrm{and~}\exists~i\in[k]:~\min_{\hat{\bs}\in\calS_m}|\hat{\bs}-\bS_i|_{\infty}>\delta\Big\}\\*
&\qquad+\Pr\Big\{\hat{\calW}_\beta(\calS_m)\neq \calW_\beta(\bS^k),~|\hat{\calW}_\beta(\calS_m)|=|\calW_\beta(\bS^k)|,\mathrm{and~}\forall~i\in[k]:~\min_{\hat{\bs}\in\calS_m}|\hat{\bs}-\bS_i|\leq\delta\Big\}\\
\nn&\leq \Pr\Big\{\exists~i\in[k]:~\min_{\hat{\bs}\in\calS_m}|\hat{\bs}-\bS_i|_{\infty}>\delta\Big\}\\*
&\qquad+\Pr\Big\{\hat{\calW}_\beta(\calS_m)\neq \calW_\beta(\bS^k),~|\hat{\calW}_\beta(\calS_m)|=|\calW_\beta(\bS^k)|,~\mathrm{and~}\forall~i\in[k]:~\min_{\hat{\bs}\in\calS_m}|\hat{\bs}-\bS_i|\leq\delta\Big\}\\
&\leq \varepsilon+\Pr\Big\{\hat{\calW}_\beta(\calS_m)\neq \calW_\beta(\bS^k),~|\hat{\calW}_\beta(\calS_m)|=|\calW_\beta(\bS^k)|~\mathrm{and~}\forall~i\in[k]:~\min_{\hat{\bs}\in\calS_m}|\hat{\bs}-\bS_i|\leq\delta\Big\}\label{use000}\\
&\leq \varepsilon+\sum_{i\in[k]}\Pr\bigg\{\bW_i^\beta(\bS^k)\notin\hat{\calW}_\beta(\calS_m)~\mathrm{and}~\min_{\hat{\bs}\in\calS_m}|\hat{\bs}-\bS_i|_{\infty}\leq\delta\bigg\}\label{usedef01}\\
&=\varepsilon+\sum_{i\in[k]}\Pr\bigg\{\bW_i^\beta(\bS^k)\notin\hat{\calW}_\beta(\calS_m)~\mathrm{and}~\exists~\hat{\bs}\in\calS_m:~|\hat{\bs}-\bS_i|_{\infty}\leq\delta\bigg\}\\
&\leq \varepsilon+\sum_{i\in[k]}\Pr\bigg\{\exists~\hat{\bs}\in\calS_m:~\bW_i^\beta(\bS^k)\neq \hat{\bw}(\hat{\bs})~\mathrm{and~}|\hat{\bs}-\bS_i|_{\infty}\leq\delta\bigg\}\label{newdef}\\
&\leq \varepsilon+\sum_{i\in[k]}\sum_{\hat{\bs}\in\calS_m}\Pr\bigg\{\bW_i^\beta(\bS^k)\neq \hat{\bw}(\hat{\bs})~\mathrm{and~}|\hat{\bs}-\bS_i|_{\infty}\leq\delta\bigg\}\\
&\leq \varepsilon+\sum_{i\in[k]}\sum_{\hat{\bs}\in\calS_m}2d\delta\tilM\label{simitoonetarget}\\
&\leq\varepsilon+2k|\calS_m|d\beta\label{usetilMagain}\\
&\leq \varepsilon+2k^2d\beta\label{uppercalSm},
\end{align}
where \eqref{use000} follows from \eqref{ktarget:step00}, \eqref{usedef01} follows since $\hat{\calW}_\beta(\calS_m)\neq \calW_\beta(\bS^k)$ and $|\hat{\calW}_\beta(\calS_m)|=|\calW_\beta(\bS^k)|$ imply that there exists $i\in[k]$ such that $\bW_i^\beta(\bS^k)\notin\hat{\calW}_\beta(\calS_m)$, \eqref{usetilMagain} follows from the definition of $\tilM$, \eqref{uppercalSm} follows since the size of $\calS_m$ is upper bounded by the number of targets $k$, and \eqref{simitoonetarget} follows since 
\begin{align}
\nn&\Pr\big\{\bW_i^\beta(\bS^k)\neq \hat{\bw}(\hat{\bs})~\mathrm{and~}|\hat{\bs}-\bS_i|_{\infty}\leq\delta\big\}\\
&=\Pr\big\{(W_{i,1}^\beta(\bS^k),\ldots,W_{i,d}^\beta(\bS^k))\neq (\hatw_1^\beta(\hat{\bs}),\ldots,\hatw_d^\beta(\hat{\bs}))\mathrm{~and~}\forall~j\in[d],~|\hats_j-S_{i,j}|\leq \delta\big\}\\
&=\Pr\big\{\exists~j\in[d]:~W_{i,j}^\beta(\bS^k)\neq \hatw_j^\beta(\hat{\bs})~\mathrm{and~}|\hats_j-S_{i,j}|\leq \delta\big\}\\
&\leq \sum_{j\in[d]}\Pr\big\{W_{i,j}^\beta(\bS^k)\neq \hatw_j^\beta(\hat{\bs}),~\mathrm{and~}|\hats_j-S_{i,j}|\leq \delta\big\}\\
&\leq 2d\delta\tilM\label{finaluse},
\end{align}
where \eqref{finaluse} bounds the probability that the quantized indices differ for two random variables which are close relative to $\delta$ (approximately equal to $\beta$ times the quantization level) and follows from \cite[Eq. (89)]{zhou2019twentyq} (see also \cite[Fig. 9]{zhou2019twentyq} for a figure illustration).

Therefore, the excess-resolution probability $\varepsilon$ satisfies
\begin{align}
\varepsilon
&\geq \Pr\{\hat{\calW}_\beta(\calS_m)\neq \calW_\beta(\bS^k)\mathrm{~and~}|\hat{\calW}_\beta(\calS_m)|=|\calW_\beta(\bS^k)|\}-2k^2d\beta\label{lbeps},\\
&=\Pr\big\{\Pi(\bW_\beta^{\uparrow}(\bS^k))\neq \Pi({\hat{\bw}}_\beta^{\uparrow}(\calS_m))\big\}-2k^2d\beta\label{macerrorp},
\end{align}
where \eqref{lbeps} follows from \eqref{uppercalSm} and \eqref{macerrorp} follows from the definitions of the vectors $\bW_\beta^{\uparrow}(\bS^k)$ and $\Pi({\hat{\bw}}_\beta^{\uparrow}(\calS_m))$ in Section \ref{sec:preconverse}.

\subsection{Connection with Data Transmission over a Point-to-Point Channel}

Since \eqref{ktarget:step00} and \eqref{macerrorp} hold for any distribution $f_{\bS}$, we specialize $f_{\bS}$ to the \emph{uniform} distribution $f_d^\rmU$ over $[0,1]^d$. This way, the target location vectors $\bS^k=(\bS_1,\ldots,\bS_k)$ are independent and for each $i\in[k]$, the $i$-th target location vector 
$\bS_i=(S_{i,1},\ldots,S_{i,d})$ is uniformly distributed over the unit cube of dimension $d$. Thus each quantized index $W_{i,j}^\beta(\bS^k)$ is uniformly distributed over the message set $[\tilM]$ and $\Gamma(\bW_i^\beta(\bS^k))$ is uniformly distributed over $[\tilM^d]$. 

Given queries $\calA^n=(\calA_1,\ldots,\calA_n)$, the noiseless answer $z_l$ to query $\calA_l$ is 
\begin{align}
z_l=\bbo(\bs^k\in\calA_l)=\bbo(\exists~i\in[k]:~\bs_i\in\calA_l).
\end{align}
When we consider the random target location vectors $\bS^k$, the induced random noiseless response $Z_l$ is then a Bernoulli random variable with parameter $1-(1-|\calA_l|)^k$. This is because $Z_l=1$ if for any $i\in[k]$, the target location vector $\bS_i$ lies in the region $\calA_l$. Considering the uniform distribution of $\bS_i$, the probability that no $\bS_i$  lies in the region $\calA_l$ for all $i\in[k]$ is $(1-|\calA_l|)^k$ and thus $\Pr\{Z_l=1\}=1-(1-|\calA_l|)^k$. Equivalently, $Z_l$ is the binary OR of $k$ independent random variables $(X_{l,1},\ldots,X_{l,k})$, each generated from the Bernoulli distribution with parameter $|\calA_l|$, i.e., $Z_l=\bbo(\exists~i\in[k];~X_{l,i}=1)$. The noisy response $Y_l$ is the output of passing $Z_l$ over the query-dependent channel $P_{Y|Z}^{f(|\calA_l|)}$. The marginal distribution $P^{\calA_l}_{Y_l}$ is thus induced by the distribution of $Z_l$ and the channel $P_{Y|Z}^{f(|\calA_l|)}$.

Bearing in mind the 20 questions estimation process for multiple targets, one concludes that $\Pr\big\{\Pi(\bW^{\uparrow}(\bS^k))\neq \Pi({\hat{\bw}}^{\uparrow}(\calS_m))\big\}$, the first term in \eqref{macerrorp}, is the error probability of transmitting a uniformly distributed message $W:=\Pi(\bW^{\uparrow}(\bS^k))\in[M^{dk_\rmp(\bS^k)}]$ with inputs $Z^n$ over a query-dependent noisy channel $P_{Y^n|Z^n}^{\calA^n}(\cdot)$ satisfying
\begin{align}
P_{Y^n|Z^n}^{\calA^n}(y^n|z^n)
&=\prod_{l\in[n]}P_{Y|Z}^{f(|\calA_l|)}(y_l|z_l)\label{conversechannel}.
\end{align}

\subsection{Final Steps}
For any target location vector $\bs^k$, the number of present targets $k_\rmp^\beta(\bs^k)$ takes values in $[k]$. Without loss of generality, we first consider $\bS^k=(\bS_1,\ldots,\bS_k)$ such that $k_\rmp(\bS^k)=t$ for some $t\in[k]$. Similarly to \cite[Proposition 4.4]{TanBook}, for for any $\kappa\in(0,1-\varepsilon-2k^2d\beta)$, we have
\begin{align}
\log \tilM^{dt}
&\leq \sup\bigg\{\psi\Big|\Pr\Bigg\{\sum_{l\in[n]}\log\frac{P_{Y|Z}^{f(\calA_l)}(Y_l|Z_l)}{P_{Y_l}^{\calA_l}(Y_l)}\leq \psi\Big\}\leq \varepsilon+2k^2d\beta+\kappa\bigg\}-\log\kappa\label{nconverse}.
\end{align}
The result in \eqref{nconverse} is slightly different from \cite[Proposition 4.4]{TanBook} since we replace $M$ and $\varepsilon$ in \cite[Proposition 4.4]{TanBook} with $\tilM^{dt}$ and $\varepsilon+2k^2d\beta$ respectively. Since \eqref{nconverse} holds for any sequence of queries $\calA^n$ and any decoding function $g:\calY^n\to\calS_m$ satisfying \eqref{ktarget:step00}, recalling the definition of $\tilM$, we have that for each $t\in[k]$,
\begin{align}
-dt\log\delta
&\leq \sup\bigg\{\psi\Big|\Pr\Big\{\sum_{l\in[n]}\log\frac{P_{Y|Z}^{f(\calA_l)}(Y_l|Z_l)}{P_{Y_l}^{\calA_l,t}(Y_l)}\leq \psi\Big\}\leq \varepsilon+2k^2d\beta+\kappa\bigg\}-\log\kappa-dt\log\beta\label{fblconverse:step:end}.
\end{align}

The proof of Theorem \ref{theorem:fbl:converse} is completed by using the definitions of the non-asymptotic fundamental limit $\delta^*(n,k,d,\varepsilon)$ in \eqref{def:delta*} and the mutual information density $\imath_{\calA_l}(z;y)$ in \eqref{def:mutualyz}, and using the result in \eqref{fblconverse:step:end}.

\section{Proof of Second-Order Asymptotics (Theorem \ref{result:second:ktarget})}
\label{proof:result:second}

\subsection{Achievability Proof}
We start with the non-asymptotic achievability bound in \eqref{ktarget:step2}. The probability terms are calculated with respect to $(P_{X_{[t]}Y}^{f(p,t)})^n$ unless otherwise stated. Note that for each $t\in[k]$,
\begin{align}
\Pr\Big\{(X^n_{[t]},Y^n)\notin\bigcap_{\calJ\subset[t]}\calD^{n,t}_\calJ(\gamma)\Big\}
&=\Pr\Big\{(X^n_{[t]},Y^n)\in\bigcup_{\calJ\subset[t]}(\calD_\calJ^{n,t}(\gamma))^\rmc\Big\}\\
&\leq \sum_{\calJ\subset[t]}\Pr\{(X^n_{[t]},Y^n)\in(\calD_\calJ^{n,t}(\gamma))^\rmc\}\\
&=\sum_{\calJ\subset[t]}\Pr\Big\{\sum_{l\in[n]}\imath_\calJ^{f(p),t}(X_{l,\calJ};Y_l)\leq (t-|\calJ|)d\log M+\gamma\Big\}\label{find:dominant}.
\end{align}

Choose $M$ and $\gamma$ such that for some $\varepsilon'\in(0,1)$,
\begin{align}
\gamma&=\frac{1}{2}\log n\label{choosegamma}\\
d\log M&=\min_{t\in[k]}\frac{nC_{\emptyset}(p,t)+\sqrt{nV_{\emptyset}(p,t)}\Phi^{-1}(\varepsilon')-\frac{1}{2}\log n}{l}\label{chooseM:ktarget}.
\end{align}
For any $t\in[k]$, it follows that
\begin{align}
dt\log M\leq nC_{\emptyset}(p,t)+\sqrt{nV_{\emptyset}(p,t)}\Phi^{-1}(\varepsilon)-\frac{1}{2}\log n\label{Mproperty}.
\end{align}
When $\calJ=\emptyset$, the Berry-Esseen theorem~\cite{berry1941accuracy,esseen1942liapounoff} implies that
\begin{align}
\nn&\Pr\Big\{\sum_{l\in[n]}\imath_{\emptyset}^{f(p),t}(X_{l,[t]};Y_l)\leq dt\log M+\gamma\Big\}\\*
&\leq \Pr\Big\{\sum_{l\in[n]}\imath_{\emptyset}^{f(p),t}(X_{l,[t]};Y_l)\leq nC_{\emptyset}(p,t)+\sqrt{nV_{\emptyset}(p,t)}\Phi^{-1}(\varepsilon')-\frac{1}{2}\log n+\gamma\Big\}\\
&\leq \varepsilon'+\frac{6T_{\emptyset}(p,t)}{(\sqrt{nV_{\emptyset}(p,t)})^3}\label{J=empty}.
\end{align}

Recall that the induced binary OR MAC satisfies the assumptions for \cite[Lemmas 1 and 2]{yavas2018random} (cf. Appendix \ref{just}). For any $t\in[2:k]$ and $\calJ\subset([t]\setminus\emptyset)$, similarly to \cite[Lemmas 1 and 2]{yavas2018random}, we have
\begin{align}
\frac{C_{\calJ}(p,t)}{|\calJ|}>\frac{C_{\emptyset}(p,t)}{t}\geq \frac{C_{\emptyset}(p,k)}{k}\label{imineq},
\end{align}
where \eqref{imineq} follows from standard calculation by noting that i) from the definition of $C_\calJ(p,t)$ in \eqref{def:cjpt}, 
\begin{align}
C_\calJ(p,t)
&=I(X_{[t]\setminus\calJ};Y|X_{\calJ}),
\end{align}
and ii) the joint distribution of $(X_1,\ldots,X_t,Y)$ in \eqref{def:pjointk} implies that $X_i$ and $X_j$ are \emph{not} conditionally independent given $Y$. For simplicity, let
\begin{align}
\kappa(t):=\min_{\calJ\subset([t]\setminus\emptyset)}\left(\frac{C_{\calJ}(p,t)}{|\calJ|}-\frac{C_{\emptyset}(p,t)}{t}\right)\label{def:kappat}.
\end{align}
Note that \eqref{imineq} ensures that $\kappa(t)>0$ for each $t\in[2:k]$.

For any $\varepsilon\in(0,1)$ and any $\calJ\subset([t]\setminus\emptyset)$, there exists a positive constant $\omega(t)$ such that
\begin{align}
\nn&\Pr\Big\{\sum_{l\in[n]}\imath_\calJ^{f(p),t}(X_{l,[t]};Y_l)\leq |\calJ|d\log M+\gamma\Big\}\\*
&=\Pr\bigg\{\sum_{l\in[n]}\imath_\calJ^{f(p),t}(X_{l,[t]};Y_l)\leq \frac{|\calJ|}{t}\Big(nC_{\emptyset}(p,t)+\sqrt{nV_{\emptyset}(p,t)}\Phi^{-1}(\varepsilon')-\frac{1}{2}\log n)\Big)+\gamma\bigg\}\label{useMprop}\\
&\leq \Pr\bigg\{\sum_{l\in[n]}\imath_\calJ^{f(p),t}(X_{l,[t]};Y_l)\leq n(C_{\calJ}(p,t)-\kappa(t))+\sqrt{nV_{\emptyset}(p,t)}\Phi^{-1}(\varepsilon')\bigg\}\label{listreasons}\\
&\leq \exp(-n \omega(t))\label{largedev},
\end{align}
where \eqref{useMprop} follows from \eqref{Mproperty}, \eqref{listreasons} follows since $1\leq |\calJ|<t$ and \eqref{def:kappat} implies that
\begin{align}
\frac{|\calJ|C_\emptyset(p,t)}{t}\leq C_\calJ(p,t)-|\calJ|\kappa(t)\leq C_\calJ(p,t)-\kappa(t),
\end{align}
and \eqref{largedev} follows from the Chernoff bound that establishes the exponential convergence of the probability term in \eqref{listreasons} with the positive real number $\omega(t)$ being the exponent.

Combining \eqref{ktarget:step2}, \eqref{find:dominant}, \eqref{J=empty} and \eqref{largedev}, when $\gamma$ and $M$ satisfy \eqref{choosegamma} and \eqref{chooseM:ktarget} respectively, we have
\begin{align}
\mathbb{E}[\rmP_\rme(\bS^k,\bX)]
&\leq 4n\exp(-2M^d\eta^2)+\exp(n\mu\eta c(f(p)))\bigg(\frac{(k+1)2^k}{\sqrt{n}}+\varepsilon'+\max_{t\in[k]}\Big(\frac{6T_{\emptyset}(p,t)}{(\sqrt{nV_{\emptyset}(p,t)})^3}+2^t\exp(-n\omega)\Big)\bigg)\\*
&\leq 4n\exp(-2M^d\eta^2)+\exp(n\mu\eta c(f(p)))\bigg(\frac{(k+1)2^k}{\sqrt{n}}+\varepsilon'+2^k\exp(-n\min_{t\in[k]}\omega(t))+\max_{t\in[k]}\frac{6T_{\emptyset}(p,t)}{(\sqrt{nV_{\emptyset}(p,t)})^3}\bigg)\label{ach:almosdtone},
\end{align}
The result in \eqref{ach:almosdtone} implies that there exist binary vectors $\tilde{\bx}=(\tilx^n(1),\ldots,\tilx^n(M^d))$ such that the excess-resolution probability with respect to the resolution level $\frac{1}{M}$ for any location vectors $\bs^k=(\bs_1,\ldots,\bs_k)$ is upper bounded by the right-hand side of \eqref{ach:almosdtone} with $M$ chosen in \eqref{chooseM:ktarget}.

Let
\begin{align}
\eta:=\sqrt{\frac{d\log M}{2M^d}}=O\left(\frac{\sqrt{n}}{\exp(nC(k)/2)}\right).
\end{align}
Similarly to \cite[Eq. (104)-(106)]{zhou2019twentyq}, we can verify that as $n\to\infty$,
\begin{align}
4n\exp(-2M^d\eta^2)=\frac{4n}{M^d}
=O\left(\frac{n}{\exp(nC(k))}\right)\to 0,
\end{align}
and
\begin{align}
\exp(n\eta cf(p))
&=1+n\eta cf(p)+o(n\eta cf(p))\\*
&=1+O\left(\frac{\sqrt{n}^3}{\exp(nC(k)/2)}\right)\to 1.
\end{align}
Furthermore, since $\calX$ and $\calY$ are both finite sets, from \cite[Lemma 47]{polyanskiy2010finite}, we conclude that the dispersion $\rmV_\emptyset(p,t)$ and the third absolute moment $T_\emptyset(p,t)$ are both finite for any $t\in[k]$. We choose $\varepsilon'$ such that the bound in \eqref{ach:almosdtone} is exactly $\varepsilon\in(0,1)$. The above analysis implies that $\varepsilon'=\varepsilon-\Theta(1/\sqrt{n})$.

Note that the above result holds for any $p\in(0,1)$. Thus, for any $n\in\bbN$ and $\varepsilon\in(0,1)$, the minimal achievable resolution $\delta^*(n,k,d,\varepsilon)$ satisfies 
\begin{align}
-\log \delta^*(n,k,d,\varepsilon)
&\geq  \max_{p\in(0,1)}\min_{t\in[k]}\frac{nC_{\emptyset}(p,t)+\sqrt{nV_{\emptyset}(p,t)}\Phi^{-1}(\varepsilon')-\frac{1}{2}\log n}{dt}\\
&=\max_{p\in(0,1)}\min_{t\in[k]}\frac{nC_{\emptyset}(p,t)+\sqrt{nV_{\emptyset}(p,t)}\Phi^{-1}(\varepsilon)-\frac{1}{2}\log n+O(1)}{dt}\label{lbbeforefinal}\\
&\geq  \max_{p\in(0,1)}\frac{nC_{\emptyset}(p,k)+\sqrt{nV_{\emptyset}(p,k)}\Phi^{-1}(\varepsilon)-\frac{1}{2}\log n+O(1)}{dk}\label{uselemma47}\\
&=\frac{C(k)+\sqrt{n\rmV(k,\varepsilon)\Phi^{-1}(\varepsilon)}-\frac{1}{2}\log n+O(1)}{dk}\label{usedefinitions},
\end{align}
where \eqref{lbbeforefinal} follows from Taylor expansion of $\rmQ^{-1}(\cdot)$, \eqref{uselemma47} follows by using \cite[Lemma 49]{polyanskiy2010thesis} (see also \cite[Appendix J]{polyanskiy2010finite}) and the second inequality in \eqref{imineq}, and \eqref{usedefinitions} follows from and the definitions of $C(k)$ in \eqref{def:maxpc} and $\rmV(k,\varepsilon)$ in \eqref{def:dispersion}.

\subsection{Converse Proof}
\label{proof:second:converse}
We start by showing that the unconditional mutual information $\imath_{\calA}(Z;Y)$ in \eqref{def:mutualyz} is equivalent to the mutual information density $\imath_\emptyset^{|\calA|,k}(X_{[k]};Y)$ in \eqref{def:cdmi} for any query $\calA\in[0,1]^d$. Recall that $X_{[k]}=(X_1,\ldots,X_k)$, each $X_i$ is a Bernoulli random variable with parameter $|\calA|$, $Z$ is a Bernoulli random variable with parameter $1-(1-f(|\calA|))^k$. Equivalently, $Z$ is the binary OR of $X_{[k]}$, i.e., $Z=\{\exists~i\in[k]:~X_i=1\}$. Thus in both cases, $Y$ is the output of passing $Z$ over the query-dependent noisy channel $P_{Y|Z}^{f(|\calA|)}$. Furthermore, from the definition of the joint distribution $P_{X_{[k]}Y}$ (cf. \eqref{def:pjointk} with $k$ replaced by $t$), the conditional probability $P_{Y|X_{[k]}}$ depends on $X_{[k]}$ only through $Z$. Therefore, both mutual information densities are equivalent since they have the same mean, variance and higher order moments.

We now proceed with the non-asymptotic converse in bound \eqref{fbl:conversebd}.
%which states that
% \begin{align}
% -\log \delta^*(n,k,d,\varepsilon)
% &\leq \max_{\calA^n\in([0,1]^d)^n}\frac{\sup\bigg\{\psi\Big|\Pr\bigg\{\sum_{l\in[n]}\imath_{\calA_l}(Z_l;Y)\leq \psi\bigg\}\leq \varepsilon+2k^2d\beta+\kappa\bigg\}-\log\kappa}{dk}.
% \end{align}
For any queries $\calA^n\in([0,1]^d)^n$, define the following first-, second- and third-order moments of the information density $\imath_{\emptyset}^{|\calA_l|,t}(X_{l,[t]};Y_l)$:
\begin{align}
C_{\calA^n}
&:=\frac{1}{n}\sum_{l\in[n]}\mathbb{E}\Big[\imath_{\calA_l}(Z_l;Y_l)\Big],\label{def:cant}\\
\rmV_{\calA^n}
&:=\frac{1}{n}\sum_{l\in[n]}\mathrm{Var}\Big[\imath_{\calA_l}(Z_l;Y_l)\Big],\\
\rmT_{\calA^n}
&:=\frac{1}{n}\sum_{l\in[n]}\mathbb{E}\bigg[\Big|\imath_{\calA_l}(Z_l;Y_l)-\mathbb{E}\Big[\imath_{\calA_l}(Z_l;Y_l)\Big]\Big|^3\bigg].
\end{align}
Since we consider finite input and output alphabets, the moments $C_{\calA^n}$, $\rmV_{\calA^n}$ and  $\rmT_{\calA^n}$ are all finite.

Choose $\gamma=\frac{1}{2}\log n$ and $\beta=\kappa=\frac{1}{\sqrt{n}}$. Consider those $\calA^n$ such that $\rmV_{\calA^n}$ is strictly positive, i.e., there exists $V_->0$ satisfying $V_-\leq \rmV_{\calA^n}$.  The Berry-Esseen theorem implies that for any $\calA^n\in([0,1]^d)^n$,
\begin{align}
\sup\Big\{\psi:~\Pr\Big\{\sum_{l\in[n]}\imath_{\calA_l}(Z_l;Y_l)\leq \psi\Big\}\leq\varepsilon+2k^2d\beta+\kappa\Big\}\Big\}\leq nC_{\calA^n}+\sqrt{n\rmV_{\calA^n}}\Phi^{-1}\bigg(\varepsilon+\frac{2k^2d+1}{\sqrt{n}}+\frac{6\rmT_{\calA^n}}{\sqrt{nV_-^3}}\bigg)\label{berryconverse1}.
\end{align}

For queries $\calA^n$ such that $\rmV_{\calA^n}=0$, the information density $\imath_{\calA_l}(Z_l;Y_l)$ is a constant for each $l\in[n]$. It follows that
\begin{align}
\sup\bigg\{\psi:~\Pr\Big\{\sum_{l\in[n]}\imath_{\calA_l}(Z_l;Y_l)\leq \psi\Big\}\leq\varepsilon+2k^2d\beta+\kappa\Big\}\bigg\}\leq nC_{\calA^n}\label{holdagain}.
\end{align}

Combining \eqref{berryconverse1} and \eqref{holdagain}, we have
\begin{align}
\nn&\frac{1}{dk}\bigg(\sup\Big\{\psi\Big|\Pr\Big\{\sum_{l\in[n]}\imath_{\calA_l}(Z_l;Y)\leq \psi\bigg\}\leq \varepsilon+2k^2d\beta+\kappa\Big\}-\log\kappa\Big)\\*
&\leq\frac{nC_{\calA^n}+\sqrt{n\rmV_{\calA^n}}\Phi^{-1}\bigg(\varepsilon+\frac{2k^2d+1}{\sqrt{n}}+\frac{6\rmT_{\calA^n}^t}{\sqrt{nV_-^3}}\bigg)+\frac{1}{2}\log n}{dk}\label{touselemma2}\\
&=:R(\calA^n,k,d).
\end{align}

Recall the definitions of $C_\emptyset(p,t)$ in \eqref{def:cemptypt}, $C(k)$ in \eqref{def:maxpc}. It follows that
\begin{align}
\sup_{\calA^n\in([0,1])^n}C_{\calA^n}
&\leq \sup_{\calA\in[0,1]^d}C_\emptyset(|\calA|,k)\label{whyhold}\\
&=\max_{p\in(0,1)}C_{\emptyset}(p,k)\\
&=C(k),
\end{align}
where \eqref{whyhold} follows since $\imath_{\calA_l}(Z_l;Y_l)$ and $\imath_\emptyset^{|\calA_l|,k}(X_{[k]};Y)$ have the same moments and thus
\begin{align}
C_{\calA^n}
&=\frac{1}{n}\sum_{l\in[n]}\mathbb{E}\Big[\imath_{\calA_l}(Z_l;Y_l)\Big]\\
&=\frac{1}{n}\sum_{l\in[n]}\mathbb{E}\Big[\imath_\emptyset^{|\calA_l|,k}(X_{[k]};Y)\Big]\\
&=\frac{1}{n}\sum_{l\in[n]}C_{\emptyset}(|\calA_l|,k)\\
&\leq \sup_{\calA\in[0,1]^d}C_\emptyset(|\calA|,k).
\end{align}

For any $p^*$ that achieves $C(k)$, it follows that
\begin{align}
\sup_{\calA^n}R(\calA^n,k,d)
&=\sup_{\calA^n:~\forall l\in[n],|\calA_l|=p^*}R(\calA^n,k,d)+O(1)\label{polydomi}\\
&\leq \sup_{\calA^n:~\forall l\in[n],|\calA_l|=p^*}\frac{1}{dk}\bigg(nC_{\calA^n}+\sqrt{n\rmV_{\calA^n}}\Phi^{-1}\bigg(\varepsilon+\frac{2k^2d+1}{\sqrt{n}}+\frac{6\rmT_{\calA^n}^k}{\sqrt{nV_-^3}}\bigg)+\frac{1}{2}\log n\bigg)+O(1)\\
&\leq \frac{nC(k)+\sqrt{n\rmV(k,\varepsilon)}\Phi^{-1}(\varepsilon)+\frac{1}{2}\log n+O(1)}{dk}\label{upptr},
\end{align}
where \eqref{polydomi} follows from \cite[Lemma 49]{polyanskiy2010thesis} and the fact that both $C_{\calA^n}$ and $\rmV_{\calA^n}$ are finite, and \eqref{upptr} follows from the Taylor expansion of $\Phi^{-1}(\cdot)$, the fact that $\rmT_{\calA^n}^k$ is finite, the definition of $\rmV(k,\varepsilon)$ in \eqref{def:dispersion} and the result that $\rmV_{\calA^n}=\rmV_\emptyset(p^*,k)$. The latter is justified as follows:
\begin{align}
\rmV_{\calA^n}
&=\frac{1}{n}\sum_{l\in[n]}\mathrm{Var}\Big[\imath_{\calA_l}(Z_l;Y_l)\Big]\\
&=\frac{1}{n}\sum_{l\in[n]}\mathrm{Var}\Big[\imath_\emptyset^{|\calA_l|,k}(X_{[k]};Y)\Big]\label{usesamemoments}\\
&=\frac{1}{n}\sum_{l\in[n]}\rmV_\emptyset(|\calA_l|,k)\label{usedefvremp}\\
&=\rmV_\emptyset(p^*,k)\label{usesize},
\end{align}
where \eqref{usesamemoments} follows since $\imath_{\calA_l}(Z_l;Y_l)$ and $\imath_\emptyset^{|\calA_l|,k}(X_{[k]};Y)$ have the same moments, \eqref{usedefvremp} follows from the definition of $\rmV_\emptyset(p,t)$ in \eqref{def:vemptypt}, and \eqref{usesize} follows since $|\calA_l|=p^*$ for all $l\in[n]$.

The converse proof of Theorem \ref{result:second:ktarget} is completed by combining \eqref{fbl:conversebd}, \eqref{berryconverse1} and \eqref{upptr}, which leads to
\begin{align}
-\log\delta^*(n,k,d,\varepsilon)
&\leq \frac{nC(k)+\sqrt{n\rmV(k,\varepsilon)}\Phi^{-1}(\varepsilon)+\frac{1}{2}\log n+O(1)}{dk}\label{converse:laststep}.
\end{align}

\section{Conclusion}
We derived bounds on the fundamental resolution limit of optimal non-adaptive 20 questions search for multiple targets over a finite dimensional unit cube. Our non-asymptotic and second-order asymptotic results in Theorems \ref{theorem:fbl:achievability} to \ref{result:second:ktarget} provided benchmarks and yielded insights into the design of non-adaptive query procedures for multiple target search. Our results were proved using the information spectrum method for a multiple access channel~\cite[Lemma 7.10.1]{han2003information}, finite blocklength information theory~\cite{polyanskiy2010finite} and inequalities for random access channel coding~\cite{yavas2018random}.

Several avenues for future work are worthwhile. It would be of interest to investigate lower complexity non-adaptive query procedures than Algorithm \ref{procedure:nonadapt:ktaget} that nonetheless achieve the benchmarks derived in this paper. It would also be of interest to study the fundamental limit of an optimal adaptive query procedure and quantify the benefit of adaptivity. To construct an adaptive query procedure that extends Algorithm \ref{procedure:nonadapt:ktaget}, one could generalize the corresponding procedures for a single target in~\cite{zhou2019twentyq,chiu2016sequential,chiu2021}. It would also be interesting to study the 20 questions estimation problem under privacy constraints~\cite{tsitsiklis2018private,xu2019optimal} and obtain the privacy-utility tradeoff~\cite{sankar2013utility,makhdoumi2013privacy,zhou2021put} for non-adaptive and adaptive query procedures.

\appendix
\subsection{Justification that the Channel Induced by \eqref{def:pjointk} Satisfies Conditions for \cite[Lemma 1]{yavas2018random}}
\label{just}

In this subsection, we show that the binary input multiple access channel $P_{Y|X_{[k]}}^{f(p),k}$ induced by \eqref{def:pjointk} satisfies the permutation-invariant, reducibility, friendliness and interference assumptions in \cite{yavas2018random}. Note that the explicit equation for $P_{Y|X_{[k]}}^{f(p),k}$ is
\begin{align}
P_{Y|X_{[k]}}^{f(p),k}(y|x_{[k]})
&:=P_{Y|Z}^{f(p)}(y|\bbo\{\exists i\in[k]:~x_j=1\})\label{virtualmac},
\end{align}
where $P_{Y|Z}^{f(p)}$ denotes the binary input point-to-point query-dependent channel. The permutation-invariant assumption states that the channel output is independent of the order of the input, i.e., $P_{Y|X_{[k]}}^{f(p),k}(y|x_{[k]})=P_{Y|X_{[k]}}^{f(p),k}(y|\hatx_{[k]})$ if $\hatx_{[k]}$ is a permutation of $x_{[k]}$. This assumption is satisfied by the channel $P_{Y|X_{[k]}}^{f(p),k}$ since as implied by \eqref{def:pjointk}, the dependence of the output $y$ on the input $x_{[k]}$ is only through whether there is a value of $1$ of the $k$ inputs $x_{[k]}$ and the order of inputs does not matter. The reducibility assumption states that if an input is zero, then it is equivalent to not having the input, i.e., for all $s<k$, $P_{Y|X_{[k]}}^{f(p),k}(y|x_{[s]},0^{k-s})=P_{Y|X_{[s]}}^{f(p),s}(y|x_{[s]})$. This assumption is clearly satisfied since the output of $y$ depends only on those inputs that have value of $1$. The friendliness assumption states that for all $s\leq k$, 
\begin{align}
I(X_{[s]};Y|X_{[s+1:k]}=0^{k-s})\geq I(X_{[s]};Y|X_{[s+1:k]})\label{toverify},
\end{align}
where the random variables $(X_{[k]},Y)\sim P_{X_{[k]}Y}^{f(p),k}$ in \eqref{def:pjointk}. To verify the result in \eqref{toverify}, note
\begin{align}
\nn&I(X_{[s]};Y|X_{[s+1:k]})\\*
&=\Pr\{X_{[s+1:k]}=0^{k-s}\}I(X_{[s]};Y|X_{[s+1:k]}=0^{k-s})+\Pr\{X_{[s+1:k]}\neq 0^{k-s}\}I(X_{[s]};Y|X_{[s+1:k]}\neq 0^{k-s})\\
&=\Pr\{X_{[s+1:k]}=0^{k-s}\}I(X_{[s]};Y|X_{[s+1:k]}=0^{k-s})\label{explain1}\\
&\leq I(X_{[s]};Y|X_{[s+1:k]}=0^{k-s}),
\end{align}
where \eqref{explain1} follows since the uncertainty of the channel output $Y$ remains unchanged if there exists $X_i=1$ in the $k$ inputs $X_{[k]}$, which implies that $H(Y|X_{[s+1:k]}\neq 0^{k-s})=H(Y|X_{[s]},X_{[s+1:k]}\neq 0^{k-s})$ and thus $I(X_{[s]};Y|X_{[s+1:k]}\neq 0^{k-s})=0$.

The interference assumption states that for any $t\in[2:k]$ and $s\in[1:t-1]$, $X_{[s]}$ and $X_{[s+1:t]}$ are conditionally dependent given $Y$, i.e., $P_{X_{[t]}|Y}\neq P_{X_{[s]}|Y}P_{X_{[s+1:t]}|Y}$. Equivalently, we need to prove that $P_{X_{[s+1:t]}|X_{[s]}Y}\neq P_{X_{[s+1]}|Y}$. For any $t\in[2:k]$ and $s\in[1:t-1]$, given the joint distribution of $(X_{[k]},Y)$ in \eqref{def:pjointk}, it follows that for any $x_{[s]}=(x_1,\ldots,x_s)$, $x_{[s+1:t]}=(x_{s+1},\ldots,x_t)$ and any $y$,
\begin{align}
P_{X_{[s]}}(x_{[s]})&=\prod_{i\in[s]}P_X(x_i)\\
P_{X_{[s+1:t]}}(x_{[s+1:t]})&=\prod_{i\in[s+1:t]}P_X(x_i)\\
P_{Y|X_{[s]}X_{[s+1:t]}}(y|x_{[s]},x_{[s+1:t]})
&=\sum_{x_{[t+1:k]}}\Big(\prod_{i\in[t+1:k]}P_X(x_i)\Big)P_{Y|X_{[k]}}(y|(x_{[s]},x_{[s+1:t]},x_{t+1:k}))\\*
P_{X_{[s]}X_{[s+1:t]}Y}(x_{[s]},x_{[s+1:t]},y)
&=P_{X_{[s]}}(x_{[s]})P_{X_{[s+1:t]}}(x_{[s+1:t]})P_{Y|X_{[s]}X_{[s+1:t]}}(y|x_{[s]},x_{[s+1:t]}).
\end{align}
All other distributions are induced by $P_{X_{[s]}X_{[s+1:t]}Y}$. In particular, the conditional distribution $P_{X_{[s+1]}|YX_{[s]}}$ satisfies
\begin{align}
P_{X_{[s+1:t]}|X_{[s]}Y}(x_{[s+1]}|x_{[s]},y)
&=\frac{P_{X_{[s]}X_{[s+1:t]}Y}(x_{[s]},x_{[s+1:t]},y)}{P_{X_{[s]}Y}(x_{[s]},y)}\\
&=\frac{P_{X_{[s]}X_{[s+1:t]}Y}(x_{[s]},x_{[s+1:t]},y)}{\sum_{x_{[s+1:t]}}P_{X_{[s]}X_{[s+1:t]}Y}(x_{[s]},x_{[s+1:t]},y)}\\
&=\frac{P_{X_{[s+1:t]}}(x_{[s+1:t]})P_{Y|X_{[s]}X_{[s+1:t]}}(y|x_{[s]},x_{[s+1:t]})}{\sum_{x_{[s+1:t]}}P_{X_{[s+1:t]}}(x_{[s+1:t]})P_{Y|X_{[s]}X_{[s+1:t]}}(y|x_{[s]},x_{[s+1:t]})}
\label{condlast}.
\end{align}
Both the numerator and the denominator in \eqref{condlast} depend on $x_{[s]}$ and the terms involving $x_{[s]}$ cannot cancelled since $P_{Y|X_{[k]}}$ depends on all elements of $X_{[k]}=(X_{[s]},X_{[s+1:t]},X_{[t+1:k]})$. Therefore, $P_{X_{[s+1:t]}|X_{[s]}Y}\neq P_{X_{[s+1:t]}|Y}$ for any $t\in[2:k]$ and $s\in[1:t-1]$ and the interference assumption is established.

\section*{Acknowledgments}
The authors would like to acknowledge five anonymous reviewers who provided many comments and suggestions that helped to significantly improve the quality of the current manuscript. The authors also acknowledge an anonymous reviewer of~\cite{zhou2019twentyq} for discussions that inspired the problem formulation of the current manuscript, especially the definition in \eqref{def:excessresolution3}. Finally, L. Zhou acknowledges Lei Yu from Nankai university for discussions on the justification of the interference assumption in Appendix \ref{just}.

\bibliographystyle{IEEEtran}
\bibliography{IEEEfull_lin}

\end{document}